\documentclass[9pt]{article}
\begin{document}
\pagestyle{plain}
\setcounter{page}{1}
\begin{center}
{\large\bf Quantum Field Theory Solution To The Gauge Hierarchy And Cosmological
Constant Problems}
\vskip 0.3 true in
{\large J. W. Moffat}
\vskip 0.3 true in {\it Department of Physics, University of Toronto,
Toronto, Ontario M5S 1A7, Canada}
\end{center}
\begin{abstract}%
A quantum field theory formalism is reviewed that leads to a self-consistent, finite
quantum gravity, Yang-Mills and Higgs theory, which is unitary and gauge invariant to
all orders of perturbation theory. The gauge hierarchy problem is solved due to the
exponential damping of the Higgs self-energy loop graph for energies greater than a
scale $\Lambda_H\leq 1$ TeV. The cosmological constant problem is solved by
introducing a fundamental quantum gravity scale, $\Lambda_G\leq 10^{-4}$ eV, above
which the virtual contributions to the vacuum energy density coupled to gravity are
exponentially suppressed, yielding an observationally acceptable value for the
particle physics contribution to the cosmological constant. Classical Einstein
gravity retains its causal behavior as well as the standard agreement with
observational data. Possible experimental tests of the onset of quantum nonlocality
at short distances are considered.
\end{abstract}



\section{Introduction}

We shall shall base our study of the hierarchy problems on
a finite quantum field theory (FQFT) which is gauge
invariant, finite and unitary to all orders of perturbation
theory [1-16]. In contrast to superstring theory and
membrane theory~\cite{Schwarz,Polchinski}, our FQFT
formalism {\it does not require the existence of extra dimensions
to guarantee its consistency}, so we shall develop our field
theory in a four-dimensional spacetime. This has the advantage
over string and membrane theory that we do not have to concern
ourselves with the problems of compactification. In string theory
there are numerous ways to compactify to lower dimensions and it
is difficult at present to justify any unique method for
achieving this compactification. Moreover, as has been learned
during the recent past, the string scale is not uniquely fixed at
the Planck scale~\cite{Witten}. String theory is not a
quantum field theory in the usual sense. All attempts to
formulate a string field theory have so far been unsuccessful.
Such theories demonstrate instabilities equivalent to those met
with in $\phi^3$ field theories, in which the Hamiltonian is
unbounded from below~\cite{Zwiebach}.

Our quantum field theory has strictly local tree graphs
and nonlocal quantum loop graphs. The FQFT gauge formalism is
applied to guarantee a self-consistent quantum gravity theory
coupled to the Yang-Mills, Higgs and spinor fields. The formalism
is free of tachyons and unphysical ghosts and satisfies unitarity
to all orders of perturbation theory. It could incorporate
supersymmetry if required, in the form of a supergravity theory,
but we shall not do so here, in order to aim for as minimal a
scheme as possible. No attempt is made to unify gauge fields
with gravity, or to extend the standard model, for we wish to
focus on the hierarchy problems.

It is commonly believed that local quantum field theory is the
only way to guarantee a consistent Poincar\'e invariant
quantum mechanics~\cite{Weinberg}. If one is willing to
give up the notion of a strictly local observable, then this
belief can be shown to be incorrect. The issue depends on the
support of the field operators and for nonlocal field theories,
it can be shown that it is impossible to construct observables
whose support is a compact set. Such theories emerge as
``quasi-local'' field theories whose local behavior only acts at
distances much larger than a certain length scale $\ell$.
Nonlocal field theories were the subject of considerable study in
the 60s and 70s, because it was thought that they could cure the
problems of non-renormalizable field theories\cite{Fainberg}.

Recently, there has been renewed
interest in nonlocal field theories in connection with string
theory, M-theory and Little String Theory
(LST)~\cite{Seiberg,Kapustin}. The LSTs are generated by
decoupling gravity and other bulk modes from five-branes. One
takes, for example, $N$ coincident five-branes and considers the
limit $g_s\rightarrow\infty, M_s\rightarrow\infty$, where $g_s$
is the string coupling at spatial infinity and
$M_s=1/\sqrt{\alpha'}$ is the string scale. It was shown by
Kapustin~\cite{Kapustin} that LSTs do not possess local
observables and that due to the exponentially increasing density
of states, the Wightman functions are not polynomially bounded.
However, he showed that the nonlocality can be accomodated by
choosing a space of test functions different from the usual
Schwartz space. We shall consider these issues in more detail
in Sect. 4.

The standard gauge symmetry of local quantum field theory is
generalized to a nonlocal transformation consisting of an
inhomogeneous term, which preserves the local, quadratic part
of the action, and a nonlocal homogeneous term, which generates a
variation of the free field action that cancels the
inhomogeneous variation of the nonlocal action. This generalized
gauge transformation is similar to the nonlocal gauge
transformation of string field theory~\cite{Zwiebach}.
The key to the success of string theory lies in the nature of
this generalized gauge invariance. Its existence guarantees the
raison d'${\hat e}$tre of gauge symmetry in quantum field theory,
which is to decouple unphysical vector and tensor quanta while
maintaining Poincar\'e invariance.

The fundamental gauge hierarchy problem is
resolved because the finite scalar Higgs self-energy loop graphs
are damped exponentially at high energies above the physical
Higgs scale $\Lambda_H$ set by the FQFT formalism and by choosing
$\Lambda_H\leq 1$ TeV. The constant $\Lambda_H$ enters
naturally, because it sets the physical non-localizable energy
scale of the Higgs particle quantum loop graphs.

The cosmological constant problem is considered to be the most
severe hierarchy problem in modern
physics~\cite{Weinberg2,Sahni,Witten2,Carroll}. The problem arises
because, in contrast to classical Newtonian gravity theory, the
Einstein gravitational Lagrangian ${\cal L}_{\rm grav}$ is not
invariant under the translation ${\cal L}_{\rm grav}\rightarrow
{\cal L}_{\rm grav}+C$, where $C$ is a constant identified with
the cosmological constant $\lambda$. Many attempts to solve this
hierarchy problem have been
made~\cite{Weinberg2,Sahni,Witten2,Carroll}, and most
recently there has been a proposal to solve the problem by
postulating a composite graviton connected to string
theory~\cite{Sundrum}. A model based on (3+1) branes and a
five-dimensional bulk has also recently been proposed\cite{Sundrum2,Witten2}.

Solving the cosmological constant problem appears to demand a
low-energy mechanism to cancel soft photon loop contributions.
How can we obtain such a mechanism in the low-energy framework
without destroying the familar successes of the standard
model?

We shall propose a quantum gravity solution to the problem
based on FQFT. We can define an effective cosmological constant
\begin{equation}
\lambda_{\rm eff}=\lambda+\lambda_{\rm vac},
\end{equation}
where $\lambda$ is the ``bare'' cosmological
constant in Einstein's classical field equations,
and $\lambda_{\rm vac}$ is the contribution that arises from the
vacuum density $\lambda_{\rm vac}=8\pi G\rho_{\rm vac}$. Already
at the standard model electroweak scale $\sim 10^2$ GeV, a
calculation of the vacuum density $\rho_{\rm vac}$ results in a
discrepancy with the observational bound
\begin{equation}
\label{vacbound} \rho_{\rm vac} < 10^{-47}\, ({\rm GeV})^4,
\end{equation}
of order $10^{55}$, resulting in a a
severe fine tuning problem, since the virtual quantum
fluctuations giving rise to $\lambda_{\rm vac}$ and the ``bare''
cosmological constant $\lambda$ must cancel to an unbelievable
degree of accuracy. If we choose the quantum gravity scale
$\Lambda_G\leq 10^{-4}$ eV, then our quantum gravity theory leads
to an exponential damping of gravitational vacuum polarisation
for $p^2 \gg \Lambda_G^2$, where $p^2$ is the square of the
Euclidean graviton momentum. This suppresses the cosmological
constant $\lambda_{\rm vac}$ below the observational bound
(\ref{vacbound}). Since the graviton tree graphs in our FQFT are
identical to the standard point like, local tree graphs of
perturbative gravity, we retain classical, causal GR and
Newtonian gravity theory, and the measured value of the
gravitational constant $G$. Only the quantum gravity loop graphs
are suppressed above energies $\leq 10^{-4}$ eV. Thus, at very
low energies or large distances, the point-like, local graviton
dominates giving rise to classical Newtonian and GR dynamics.

The scales $\Lambda_H$ and $\Lambda_G$ are determined by the quantum
non-localizable nature of the Higgs particle and the
gauge particles $W$ and $Z$ of the standard model as compared to the
graviton. The Higgs particle radiative corrections have a nonlocal scale
at $\ell_H\sim 10^{-16} $ cm, whereas the graviton radiative corrections
are localizable down to a large length scale $\ell_G \leq 1$ cm. Thus, the
fundamental energy scales in the theory are determined by the underlying
physical nature of the particles and fields and do not correspond
to arbitrary cut-offs, which destroy the gauge invariances of the
field theory. The underlying explanation of these physical scales
must be sought in a more fundamental theory.

In Section 2, we describe the basic local action of the theory
and in Section 3, we provide a review of FQFT as a perturbative
quantum field theory. The nonlocal quantum behavior of the theory
is considered in detail in Section 4, and in Section 5, we discuss a possible
exerimental test of the onset of nonlocality by detecting CPT asymmetries.
In Sections 6 and 7, we
develop the formalism for Yang-Mills gauge theory and quantum
gravity. In Section 8, we turn our attention to the resolution of
the gauge hierarchy problem in the Higgs sector, while in
Section 9, we analyze the results of gluon and
gravitational vacuum polarization calculations. In Section 10, we
use FQFT quantum gravity to resolve the cosmological constant problem and
in Section 11, we end with concluding remarks.

\section{\bf The Action}

We shall begin with the four-dimensional action
\begin{equation}
W=W_{\rm grav}+W_{YM}+W_{\rm H}+W_{\rm Dirac}+W_M,
\end{equation}
where
\begin{equation}
W_{\rm grav}=-\frac{2}{\kappa^2}\int d^4x\sqrt{-g}(R+2\lambda),
\end{equation}
\begin{equation}
W_{\rm YM}=-\frac{1}{4}\int d^4x\sqrt{-g}\,{\rm Tr}(F^2),
\end{equation}
\begin{equation}
W_{\rm H}=-\frac{1}{2}\int
d^4x\sqrt{-g}[D_\mu\phi^iD^\mu\phi^i+V(\phi^2)],
\end{equation}
\begin{equation}
W_{\rm Dirac}=\frac{1}{2}\int
d^4x\sqrt{-g}\bar{\psi}\gamma^ae_a^\mu[\partial_\mu\psi-\omega_\mu\psi
-{\cal D}(A_{i\mu})\psi]+h.c.
\end{equation}
Here, we use the
notation: $\mu,\nu=0,1,2,3$, $g={\rm det}(g_{\mu\nu})$ and the
metric signature of Minkowski spacetime is $\eta_{\mu\nu}={\rm
diag}(-1,+1,+1,+1)$. The Riemann tensor is defined such that
\begin{equation}
{R^\lambda}_{\mu\nu\rho}=\partial_\rho{\Gamma_{\mu\nu}}^\lambda
-\partial_\nu{\Gamma_{\mu\rho}}^\lambda+
{\Gamma_{\mu\nu}}^\alpha{\Gamma_{\rho\alpha}}^\lambda
-{\Gamma_{\mu\rho}}^\alpha{\Gamma_{\nu\alpha}}^\lambda.
\end{equation}
Moreover, h.c. denotes the Hermitian conjugate,
$\bar\psi=\psi^{\dagger} \gamma^0$, and $e^\mu_a$ is a vierbein,
related to the metric by
\begin{equation}
g_{\mu\nu}=\eta_{ab}e_\mu^ae_\nu^b,
\end{equation} where
$\eta_{ab}$ is the four-dimensional Minkowski metric tensor
associated with the flat tangent space with indices a,b,c...
Moreover, $F^2= F_{i\mu\nu}F^{i\mu\nu}$, $R$ denotes the scalar
curvature, $\lambda$ is the cosmological constant and
\begin{equation} F_{i\mu\nu}=\partial_\nu A_{i\mu}-\partial_\mu
A_{i\nu}-ef_{ikl}A_{k\mu}A_{l\nu},
\end{equation} where
$A_{i\mu}$ are the gauge fields of the Yang-Mills group with
generators $f_{ikl}$, $e$ is the coupling constant and
$\kappa^2=32\pi G$ with $c=1$. We denote by $D_\mu$ the
covariant derivative operator
\begin{equation}
D_\mu\phi^i=\partial_\mu\phi^i+ef^{ikl}A_\mu^k\phi^l.
\end{equation}
The Higgs potential $V(\phi^2)$ is of the form
leading to spontaneous symmetry breaking
\begin{equation}
V(\phi^2)=\frac{1}{4}g(\phi^i\phi^i-K^2)^2+V_0,
\end{equation}
where $V_0$ is an adjustable constant and the coupling constant
$g > 0$.

The spinor field is minimally coupled to the gauge potential
$A_{i\mu}$, and ${\cal D}$ is a matrix representation of the
gauge group $SO(3,1)$. The spin connection $\omega_\mu$ is
\begin{equation}
\omega_\mu=\frac{1}{2}\omega_{\mu ab}\Sigma^{ab},
\end{equation} where
$\Sigma^{ab}=\frac{1}{4}[\gamma^a,\gamma^b]$ is the spinor matrix
associated with the Lorentz algebra $SO(3,1)$. The components
$\omega_{\mu ab}$ satisfy
\begin{equation}
\partial_\mu
e^\sigma_a+{\Gamma_{\mu\nu}}^\sigma e^\nu_a-{\omega_{\mu a}}^\rho
e_\rho^\sigma=0,
\end{equation} where ${\Gamma_{\mu\nu}}^\sigma$
is the Christoffel symbol. The field equations for the
gravity-Yang-Mills-Higgs-Dirac sector are
\begin{equation}
R_{\mu\nu}-\frac{1}{2}g_{\mu\nu}R-\lambda
g_{\mu\nu}=-\frac{1}{4}\kappa^2T_{\mu\nu},
\end{equation}
\begin{equation}
g^{\rho\mu}\nabla_\rho
F^i_{\mu\nu}=g^{\rho\mu}\biggl(\partial_\rho
F^i_{\mu\nu}-{\Gamma_{\rho\mu}}^\sigma F^i_{\sigma\nu}
$$ $$
-{\Gamma_{\rho\nu}}^\sigma F^i_{\mu\sigma}
+[A_\rho,F_{\mu\nu}]^i\biggr)=0,
\end{equation}
\begin{equation}
\frac{1}{\sqrt{-g}}D_\mu[\sqrt{-g}g^{\mu\nu}D_\nu\phi^i]
=\biggl(\frac{\partial V}{\partial\phi^2}\biggr)\phi^i,
\end{equation}
\begin{equation}
\gamma^ae^\mu_a[\partial_\mu-\omega_\mu-{\cal D}(A_\mu)]\psi=0.
\end{equation}
The energy-momentum tensor is
\begin{equation}
T_{\mu\nu}=T^{\rm YMH}_{\mu\nu}+T^{\rm Dirac}_{\mu\nu}+T^{\rm
M}_{\mu\nu},
\end{equation}
where
\begin{equation}
T_{\mu\nu}^{\rm YMH}={\rm Tr}(F_{\mu\sigma}
F^\sigma_\nu)+D_\mu\phi^iD_\nu\phi^i
$$ $$
-\frac{1}{2}g_{\mu\nu}\biggl[\frac{1}{2}{\rm Tr}(F^2)
+D_\sigma\phi^iD^\sigma\phi^i+V(\phi^2)\biggr],
\end{equation}
\begin{equation}
T^{\rm Dirac}_{\mu\nu}
=-\bar\psi\gamma_\mu[\partial_\nu-\omega_\nu
-{\cal D}(A_{i\nu})]\psi,
\end{equation}
and $T_{\mu\nu}^{\rm M}$ is the energy-momentum tensor of
non-field matter.

\section{\bf Finite Quantum Field Theory Formalism}

An important development in nonlocal FQFT was the discovery that
gauge invariance and unitarity can be restored by adding series
of higher interactions. The resulting theory possesses a
nonlinear, field representation dependent gauge invariance which
agrees with the original local symmetry on shell but is larger
off shell. Quantization is performed in the functional formalism
using an analytic and convergent measure factor which retains
invariance under the new symmetry. An explicit calculation was
made of the measure factor in QED~\cite{Moffat2}, and it
was obtained to lowest order in Yang-Mills
theory~\cite{Kleppe2}. Kleppe and
Woodard~\cite{Woodard2} obtained an ansatz based on the
derived dimensionally regulated result when
$\Lambda\rightarrow\infty$, which was conjectured to lead to a
general functional measure factor in FQFT gauge theories.

In contrast to string theory, we can achieve a {\it genuine
quantum field theory}, which allows vertex operators to be taken off the
mass shell. The finiteness draws from the fact that factors of $\exp[{\cal
K}(p^2)/2\Lambda^2]$ are attached to propagators which suppress any
ultraviolet divergences in Euclidean momentum space, where $\Lambda$ is an
energy scale factor. An important feature of FQFT is {\it that only the
quantum loop graphs have nonlocal properties}; the classical tree graph
theory retains full causal and local behavior.
 
A convenient formalism which makes the FQFT construction transparent is
based on shadow fields~\cite{Kleppe2,Woodard2}. We shall consider
the 4-dimensional spacetime to be approximately flat Minkowski
spacetime. Let us denote by $f_i$ a generic local field and write
the standard local action as
\begin{equation}
W[f]=W_F[f]+W_I[f],
\end{equation}
where $W_F$ and $W_I$ denote the free part and the interaction part
of the action, respectively, and
\begin{equation}
W_F[f]=\frac{1}{2}\int d^4xf_i{\cal K}_{ij}f_j.
\end{equation}
In a gauge theory $W$ would be the Becchi, Rouet, Stora, Tyutin (BRST)
gauge-fixed action including ghost fields in the invariant action required
to fix the gauge\cite{Becchi}. The kinetic operator ${\cal K}$ is fixed by
defining a Lorentz-invariant distribution operator
\begin{equation}
\label{distribution}
{\cal E}\equiv \exp\biggl(\frac{{\cal K}}{2\Lambda^2}\biggr)
\end{equation} and the shadow operator: \begin{equation}
{\cal O}^{-1}=\frac{{\cal K}}{{\cal E}^2-1}.
\end{equation}

Every local field $f_i$ has an auxiliary counterpart field $h_i$, and they
are used to form a new action
\begin{equation}
W[f,h]\equiv W_F[\hat f]-P[h]+W_I[f+h],
\end{equation}
where
\[
\hat f={\cal E}^{-1}f,\quad P[h]=\frac{1}{2}\int d^4xh_i{\cal
O}^{-1}_{ij} h_j.
\]
By iterating the equation
\begin{equation}
\label{iteration}
h_i={\cal O}_{ij}\frac{\delta W_I[f+h]}{\delta h_j}
\end{equation}
the shadow fields can be determined as functions, and the regulated
action is derived from
\begin{equation}
\hat W[f]=W[f,h(f)].
\end{equation}
We recover the original local action when we take the limit
$\Lambda\rightarrow\infty$ and $\hat f\rightarrow f, h(f)\rightarrow 0$.

The expression (\ref{iteration}) can be developed into a series
expansion for $h_i[f]$. The regularized action is found by
substituting into it the classical solution $h_i[f]$. Expanding
${\hat W}$ in powers of $f$ gives the kinetic term $W_F[{\hat
f}]$, together with an infinite series of interaction terms the
first of which is just $W_I[f]$. Since ${\cal O}$ is an entire
function of ${\cal K}$ the higher interactions are also entire
functions of ${\cal K}$. This is important for preserving
unitarity.

Quantization is performed using the definition
\begin{equation}
\langle 0\vert T^*(O[f])\vert 0\rangle_{\cal E}=\int[Df]\mu[f]({\rm gauge\,
fixing})O[\hat f]\exp(i\hat W[f]).
\end{equation}
On the left-hand side we have the regulated vacuum expectation value of the
$T^*$-ordered product of an arbitrary operator $O[f]$ formed from the local
fields $f_i$. The subscript ${\cal E}$ signifies that a regulating Lorentz
distribution has been used. Moreover, $\mu[f]$ is a measure factor and
there is a gauge fixing factor, both of which are needed to maintain
perturbative unitarity in gauge theories.

The new Feynman rules for FQFT are obtained as follows: The vertices remain
unchanged within the regularized action, but every leg of a
diagram is connected either to a regularized propagator,
\begin{equation}
\label{regpropagator}
\frac{i{\cal E}^2}{{\cal K}+i\epsilon}
=-i\int^{\infty}_1\frac{d\tau}{\Lambda^2}\exp\biggl(\tau
\frac{{\cal K}}{\Lambda^2}\biggr),
\end{equation}
or to a shadow propagator,
\begin{equation}
-i{\cal O}=\frac{i(1-{\cal E}^2)}{{\cal K}}=-i\int^1_0\frac{d\tau}
{\Lambda^2}
\exp\biggl(\tau\frac{{\cal K}}{\Lambda^2}\biggr).
\end{equation}
We shall also attach a factor ${\cal E}(p^2)$ to every external
leg connected to a loop, which is unity on shell. The formalism
is set up in Minkowski spacetime and loop integrals are formally
defined in Euclidean space by performing a Wick rotation. This
facilitates the analytic continuation; the whole formalism could
from the outset be developed in Euclidean space.

In FQFT renormalization is carried out as in any other field theory.
The bare parameters are calculated from the renormalized ones and
$\Lambda$, such that the limit $\Lambda\rightarrow\infty$ is finite for
all noncoincident Green's functions, and the bare parameters are those of
the local theory. The regularizing interactions {\it are determined by the
local operators.}

The regulating Lorentz distribution function ${\cal E}$ must be chosen to
perform an explicit calculation in perturbation theory. We do not know the
unique choice of ${\cal E}$. However, once a choice for the function is made, then the
theory and the perturbative calculations are uniquely fixed. A standard
choice in early FQFT papers is~\cite{Moffat,Moffat2}:
\begin{equation}
\label{reg}
{\cal E}_m=\exp\biggl(\frac{\partial^2-m^2}{2\Lambda^2}\biggr).
\end{equation}

An explicit construction for QED was given using the Cutkosky rules as
applied to FQFT whose propagators have poles only where ${\cal K}=0$ and
whose vertices are entire functions of ${\cal K}$. The regulated action
$\hat W[f]$ satisfies these requirements which guarantees unitarity on the
physical space of states. The local action is gauge fixed and then a
regularization is performed on the BRST theory.

The infinitesimal transformation
\begin{equation}
\delta f_i=T_i(f)
\end{equation}
generates a symmetry of $W[f]$, and the infinitesimal
transformation
\begin{equation}
\label{nonloctran}
\hat\delta f_i={\cal E}^2_{ij}T_j(f+h[f])
\end{equation}
generates a symmetry of the regulated action ${\hat W}[f]$. To
see this consider the transformations
\begin{equation}
\label{transf}
\delta f_i={\cal E}^2_{ij}T_j[f+h],\quad
\delta h_i=(1-{\cal E}^2)_{ij}T_j[f+h].
\end{equation}
Adding these two transformations gives
\begin{equation}
\delta(f+h)_i=T_i[f+h].
\end{equation}
Then, (\ref{transf}) is a symmetry of the action $W[f,h]$. We
have
\begin{equation}
\delta W[f,h]=\int
d^4x\biggl\{(f_i+h_i){\cal K}_{ij}T_j[f+h]\biggr\} $$ $$
+\frac{\delta W_I[f+h]}{\delta f_i}T_i[f+h]
=\delta W[f+h].
\end{equation}
It follows that $\delta W[f,h]=0$ is a consequence of the assumed
invariance $\delta W[f+h]=0$. Now ${\hat W}[f]$ is invariant
under (\ref{nonloctran}), for we have
\[
{\hat\delta}h_i[f]=
(1-{\cal E}^2_{ij})T_j[f+h[f]]-L_{ij}[f+h[f]]\frac{\delta T_j}
{\delta f_k}[f+h[f]]{\cal E}^2_{kl}
\frac{\delta {\hat W}[f]}{\delta f_l},
\]
where
\[
L^{-1}_{ij}={\cal O}^{-1}_{ij}-\frac{\delta^2W_I[f]}{\delta
f_i\delta f_j}.
\]

It follows that FQFT
regularization preserves all continuous symmetries including
supersymmetry. The quantum theory will preserve symmetries
provided a suitable measure factor can be found such that
\begin{equation}
\hat\delta([Df]\mu[f])=0.
\end{equation}
Moreover, the interaction vertices of the measure factor must be
entire functions of the operator ${\cal K}$ and they must not destroy the
FQFT finiteness.

In FQFT tree order, Green's functions remain local except for
external lines which are unity on shell. It follows immediately that since
on shell tree amplitudes are unchanged by the regularization, $\hat W$
preserves all symmetries of $W$ on shell. Also all loops contain at least
one regularizing propagator and therefore are ultraviolet finite. Shadow
fields are eliminated at the classical level, for functionally integrating
over them would produce divergences from shadow loops. Since shadow field
propagators do not contain any poles there is no need to quantize the
shadow fields.

In FQFT, the on shell tree amplitudes agree with the local,
unregulated action, while the loop amplitudes disagree. This
seems to contradict the Feynman tree
theorem~\cite{Feynman3}, which states that loop
amplitudes of local field theory can be expressed as sums of
integrals of tree diagrams. If two local theories agree at
the tree level, then the loop amplitudes agree as well.
However, the tree theorem does not apply to nonlocal field
theories. The tree theorem is proved by using the propagator relation
\begin{equation}
D_F=D_R+D^+
\end{equation}
to expand the Feynman propagator $D_F$ into a series in the on
shell propagator $D^+$. This decomposes all terms with even one
$D^+$ into trees. The term with no $D^+s$ is a loop formed with
the retarded propagator and vanishes for local interactions. But
for nonlocal interactions, this term generally survives and new
physical effects occur in loop amplitudes, which cannot be
predicted from the local on shell tree graphs.

\section{\bf Quantum Nonlocal Behavior in FQFT}

It appears on general grounds that interacting strings are
nonlocal~\cite{Uglum,Uglum2,Thorlacius}. Nonlocality in open
string theory can arise from the non-commutativity of spacetime
coordinates~\cite{Seiberg,Seiberg2,Seiberg3}
\begin{equation}
[x^\mu, x^\nu]=i\theta\epsilon^{\mu\nu}.
\end{equation}
This nonlocality in string theory is closely associated with the
string uncertainty principle
\begin{equation}
\Delta x\Delta t \geq \alpha'.
\end{equation}
Nonlocality has also been associated
with the formation of black hole horizons and the lack of
commutativity of spatial coordinates and
time~\cite{thooft2}. The horizon responds to incoming matter
before it comes in.

Kapustin~\cite{Kapustin} has recently shown
that LSTs are quasi-local field theories whose infrared limit can
approach local field theories in the large~\cite{Fainberg}. The
exponential growth of Wightman functions (Green's functions) in
momentum space is a characteristic feature of nonlocal field
theories. The corresponding test functions in x-space are real
analytic and cannot possess compact support.

The Wightman functions~\cite{Wightman} or vacuum
expectations values of products of field operators $\phi(p)$:
\begin{equation}
W_n(q_1,...,q_{n-1})=\langle 0
\vert\phi(q_1)\phi(q_2)...\phi(q_{n-1}) \vert 0\rangle,
\end{equation}
grow exponentially with momenta for nonlocal field theories. By
the positivity of energy, $W_n$ vanishes when any of its
arguments are outside the forward light cone. Inside the forward
light cone $W_n$ is bounded by
\begin{equation}
\exp[\ell(\vert q_1\vert+...\vert q_{n-1}\vert)],
\end{equation}
where $\vert q\vert=\sqrt{q^2}$ and $\ell$ is a length scale.
In the case of LST models, the length scale is given by $\ell\sim
\sqrt{N}/M_s$ where $N$ is the number of coincident five-branes.

Jaffe~\cite{Jaffe} defined a test function space ${\tilde
S}_g$ in momentum space, which is convenient to use when
discussing nonlocal field theories, in which all functions are
infinitely differentiable and for which all the norms are finite.
Given a positive function $g(t)$ which is entire, Jaffe showed
that if $g(t)$ satisfies
\begin{equation}
\label{Jaffe}
\int^\infty_0dt\frac{\ln g(t^2)}{1+t^2} < \infty,
\end{equation}
then the Fourier transform of ${\tilde S}_g$ has functions
with compact support, strictly local quantum fields can be
defined and a local quantum field theory can be formulated.
On the other hand, if (\ref{Jaffe}) is not satisfied, then there
are no test functions with compact support and we have a
nonlocal quantum field theory.

Our choice of the entire function ${\cal E}(p^2)$ in the factor,
Eq. (\ref{distribution}), will not lead to a test function space
that satisfies the condition (\ref{Jaffe}). We can choose a
function ${\cal E}(p^2)$ which will provide a test function space
that leads to a quasi-local quantum field theory, as defined by
Kapustin, and in the earlier work by Iofa and Fainberg. In the
present work, we have chosen ${\cal K}(p^2)=-(p^2+m^2)$, because
it leads to a simplification of calculations in perturbation
theory. But this is purely a technical issue, and we can
certainly adopt entire functions ${\cal K}(p^2)$ which lead to
quasi-local field operators, which only violate locality at short
distances.

The commutator for a scalar field operator $\phi(x)$:
\begin{equation}
[\phi(x),\phi(y)]=W_2(x-y)-W_2(y-x)
\end{equation}
in our theory will not vanish outside the light cone for
space-like separations $(x-y)^2 > 0$. Indeed, it will satisfy
\begin{equation}
[\phi(x),\phi(y)]\sim \delta[(x-y)^2 -\ell^2]{\rm sign}(x_0-y_0).
\end{equation}
In FQFT, it can be argued that the extended objects that
replace point particles (the latter are obtained in the limit
$\Lambda\rightarrow\infty$) cannot be probed because of a
Heisenberg uncertainty type of argument. The FQFT nonlocality
{\it only occurs at the quantum loop level}, so there is no
noncausal classical behavior. In FQFT the strength of a signal
propagated over an invariant interval $\ell^2$ outside the light
cone would be suppressed by a factor $\exp(-\ell^2\Lambda^2)$.

Nonlocal field theories can possess non-perturbative
instabilities. These instabilities arise because of extra
canonical degrees of freedom associated with higher time
derivatives. If a Lagrangian contains up to $N$ time derivatives,
then the associated Hamiltonian is linear in $N-1$ of the
corresponding canonical variables and extra canonical degrees of
freedom will be generated by the higher time derivatives. A
nonlocal theory can be viewed as the limit $N\rightarrow\infty$
of an Nth derivative Lagrangian. Unless the dependence on the
extra solutions is arbitrarily choppy in the limit, then the
higher derivative limit will produce
instabilities~\cite{Eliezer}. The condition for the
smoothness of the extra solutions is that no invertible field
redefinition exists which maps the nonlocal field equations into
the local ones. String theory does satisfy this smoothness
condition as can be seen by inspection of the S-matrix tree
graphs. In FQFT the tree amplitudes agree with those of the local
theory, so the smoothness condition is not obeyed.

It was proved by Kleppe and Woodard~\cite{Kleppe2} that the solutions
of the nonlocal field equations in FQFT are in one-to-one correspondence
with those of the original local theory. The relation for a generic field
$v_i$ is \begin{equation}
v_i^{\rm nonlocal}={\cal E}^2_{ij}v^{\rm local}_j.
\end{equation}
Also the actions satisfy
\begin{equation}
W[v]={\hat W}[{\cal E}^2v].
\end{equation}
Thus, there are no extra classical solutions.
The solutions of the regularized nonlocal Euler-Lagrange equations are
in one-to-one correspondence with those of the local action. It follows
{\it that the regularized nonlocal FQFT is free of higher derivative
solutions, so FQFT can be a stable theory.}

Since only the quantum loop graphs in the nonlocal FQFT differ from
the local field theory, then FQFT can be viewed as a non-canonical
quantization of fields which obey the local equations of motion. Provided
the functional quantization in FQFT is successful, then the theory does
maintain perturbative unitarity.

\section{\bf Experimental Tests of Nonlocality}

In order to solve the Higgs and cosmological constant radiative stability, hierarchy
problems, we have relaxed the assumption of microcausal locality in our FQFT. A
scale of nonlocality $\Lambda$ is set for the graviton ($\Lambda_{G}\leq
10^{-3}$ eV), the Higgs particle ($\Lambda_H\leq 1$ TeV) and the standard model gauge
particles ($\Lambda_{GP} \gg 1$ TeV). We do not understand the fundamental
physics which is the source of these nonlocality scales but, as we shall see, given
these scales we can potentially solve the radiative stability problems in a fully
gauge invariant, finite and unitary fashion, including the gravitational stability of
the cosmological constant.

Supersymmetry and technicolor models have been proposed to solve the Higgs gauge
hierarchy problem. The mass scales for supersymmetry are set 'by hand', so to speak,
according to when we expect supersymmetry breaking to set in, allowing
super-partners to be detected, and when technicolor fermions form condensates,
allowing us to detect technicolor particles. No known fundamental physics tells us
what these mass scales are. We can only guess their magnitude above certain obvious
intermediate energy bounds. Experiments already tend to disfavour techniclor models,
and if the large hadron colliders do not detect super-partners below 2-3 TeV, then
this would kill the possibility of using supersymmetric models to explain the
radiative stability of the Higgs particle.

Can we experimentally detect the onset of nonlocality? We could do this by checking
dispersion relations for scattering amplitudes at high energies. We expect that the
non-vanishing of commutators of field operators outside the light cone will decrease
exponentially with the spacelike distance, so violations of nonlocality will be
small, and changes of analyticity of the scattering amplitudes from the standard
microcausal analyticity properties will correspondingly be small. Another possible
signature of nonlocality is a violation of CPT invariance. This is a fundamental
theorem of local quantum field theory~\cite{Luders,Wightman}. There have been
suggestions that CPT invariance could be broken in quantum gravity~\cite{Ellis}.
Moreover, there have been several studies of meson decays with the prospects of
detecting CPT invariance breaking at K-meson and B-meson factories~\cite{Kmeson}.

Let us investigate how CPT invariance could be violated by nonlocality. Consider a
complex, nonlocal Heisenberg-picture scalar field operator $\Phi(x)$. The
K\"allen-Lehmann representation is given by the vacuum expectation
value~\cite{Lehmann,Weinberg}
\begin{equation} \langle
0\vert\Phi(x)\Phi^\dagger(y)\vert 0\rangle =\int_0^\infty
d\mu^2\rho(\mu^2){\tilde\Delta}_{+}(x-y;\mu^2),
\end{equation}
where
\begin{equation}
{\tilde\Delta}_{+}(x-y;\mu^2)=\frac{1}{(2\pi)^3}\int_0^\infty
d^4p\exp[ip\cdot(x-y)]\Pi(x-y)\theta(p^0)\delta(p^2+\mu^2),
\end{equation}
and $\Pi(x-y)$ is an entire analytic function with $\Pi(x) >0$ for real
$x$~\cite{Moffat}. The spectral function $\rho$ is defined by
\begin{equation}
\sum_n\delta^4(p-p_n)\vert\langle 0\vert\Phi(0)\vert n\rangle\vert^2
=\frac{1}{(2\pi)^3}\theta(p^0)\rho(-p^2)
\end{equation}
with $\rho(-p^2)=0$ for $p^2 < 0$. We also have
\begin{equation}
\langle 0\vert\Phi^\dagger(y)\Phi(x)\vert
0\rangle =\int_0^\infty d\mu^2{\bar\rho}(\mu^2){\tilde\Delta}_{+}(y-x;\mu^2),
\end{equation}
where
\begin{equation}
\sum_n\delta^4(p-p_n)\vert\langle n\vert\Phi^\dagger(0)\vert
0\rangle\vert^2 =\frac{1}{(2\pi)^3}\theta(p^0){\bar\rho}(-p^2).
\end{equation}

Let us define
\begin{equation}
\alpha(\mu^2)=\rho(\mu^2)-{\bar\rho}(\mu^2).
\end{equation}
The vacuum expectation value of the commutator is
\begin{equation}
\label{commutator}
\langle 0\vert[\Phi(x),\Phi^\dagger(y)]\vert 0\rangle
$$
$$
=\int_0^\infty d\mu^2\{\rho(\mu^2)[{\tilde\Delta}_{+}(x-y;\mu^2)-
{\tilde\Delta}_{+}(y-x;\mu^2)]+\alpha(\mu^2){\tilde\Delta}_{+}(y-x;\mu^2)\}.
\end{equation}
For spacelike separations $(x-y)^2>0$, the function ${\tilde\Delta}_{+}(x-y;\mu^2)=
{\tilde\Delta}_{+}(y-x;\mu^2)$ and it does not vanish. For (\ref{commutator}) to
vanish for spacelike separations, we must have $\alpha(\mu^2)=0$. This is a
nonperturbative proof of the CPT theorem, for states with $p^2=-\mu^2$ have the
quantum numbers of the particle associated with $\Phi$, and there must be
corresponding states with $p^2=-\mu^2$ that have the quantum numbers of the
anti-particle described by the operator $\Phi^\dagger$~\cite{Weinberg}. For strictly
local field operators $\Phi$, the commutator
\begin{equation}
[\Phi(x),\Phi(y)]=0
\end{equation}
for spacelike separations $(x-y)^2 > 0$. However, we assumed that the
$\Phi(x)$ were nonlocal field operators, so there will be a violation of the CPT
theorem when $\alpha(\mu^2)\not= 0$, and we have for spacelike separation
\begin{equation}
\langle 0\vert[\Phi(x),\Phi^\dagger(y)]\vert 0\rangle
=\int_0^\infty d\mu^2\alpha(\mu^2){\tilde\Delta}_{+}(y-x;\mu^2).
\end{equation}

The vacuum expectation value of the time-ordered product is
\begin{equation}
\langle 0\vert T\biggl\{\Phi(x)\Phi^\dagger(y)\biggr\}\vert 0\rangle
$$
$$
=i\int_0^\infty d\mu^2\rho(\mu^2){\tilde\Delta}_F(x-y;\mu^2)+i\int_0^\infty
d\mu^2\alpha(\mu^2)\theta(y^0-x^0){\tilde\Delta}_{+}(y-x;\mu^2),
\end{equation}
where $\Delta_F$ is the Feynman propagator
\begin{equation}
-i{\tilde\Delta}_F(x-y;\mu^2)=\theta(x^0-y^0){\tilde\Delta}_{+}(x-y;\mu^2)
-\theta(y^0-x^0){\tilde\Delta}_{+}(y-x;\mu^2).
\end{equation}

For a nonlocal interaction
\begin{equation}
V_{\rm NL}=\int d^3x{\cal H}_{\rm NL}(\vec{x},0),
\end{equation}
the commutator $[CPT,V_{\rm NL}]$ will not in general vanish. The masses and decay
rates of particles and anti-particles will not be equal for CPT invariance violating
processes. For the discrete symmetries of nature, violations have been observed for
C, P and the combined CP symmetries. Two types of CP symmetry violation have been
observed for K-mesons. An active pursuit to detect CPT asymmetries in meson decays is
presently underway.

\section{\bf Finite Quantum Yang-Mills Theory}

Let us now review the finite quantization of the
Yang-Mills sector in four-dimensional Minkowski flat space. The
gauge field strength $F_{i\mu\nu}$ is invariant under the
familiar transformations:
\begin{equation}
\delta
A_{i\mu}=-\partial_\mu\theta_i+ ef_{ikl}A_{k\mu}\theta_l.
\end{equation}

To regularize the Yang-Mills sector, we identify the kinetic operator
\[
{\cal K}_{ik}^{\mu\nu}
=\delta_{ik}(\partial^2\eta^{\mu\nu}-\partial^\mu\partial^\nu).
\]
The regularized action is given by\cite{Kleppe2}
\begin{equation}
\hat{W}_{YM}[A]=\frac{1}{2}\int d^4x\biggl\{\hat{A}_{i\mu}
{\cal K}^{\mu\nu}_{ik}\hat{A}_{k\nu}-B_{i\mu}[A]({\cal
O}^{\mu\nu}_{ik})^{-1} B_{k\nu}[A]\biggr\}
$$
$$
+W^I_{YM}[A+B[A]],
\end{equation}
where $B_{i\mu}$ is the Yang-Mills shadow field, which satisfies
the expansion
\begin{equation}
B^\mu_i[A]={\cal O}^{\mu\nu}_{ik}\frac{\delta
W^I_{YM}[A+B]}{\delta B^\nu_k}
$$ $$
={\cal O}^{\mu\nu}_{ik}ef_{klm}[A_{\nu l}
\partial_\sigma A^\sigma_m+A_{l\sigma}\partial_\nu A^\sigma_m
-2A_{l\sigma}\partial^\sigma A_{\nu_m}]+O(e^2A^3).
\end{equation}
The regularized gauge symmetry transformation is
\[
\hat{\delta}_{\theta}A^\mu_i=({\cal E}^{2\mu\nu}_{ik})
\biggl\{-\partial_\nu\theta_k+ef_{klm}(A_{l\nu}+B_{l\nu}[A])\theta_m\biggr\}.
\]
The extended gauge transformation is neither linear nor local.

We functionally quantize the Yang-Mills sector using
\begin{equation}
\langle 0\vert T^*(O[A])\vert 0\rangle_{\cal E}=\int[DA]
\mu[A]({\rm gauge\, fixing})
O[{\hat A}]\exp(i\hat W_{\rm YM}[A]).
\end{equation}

To fix the gauge we use Becchi-Rouet-Stora-Tyutin
(BRST)~\cite{Becchi} invariance. The ghost structure of
the BRST action comes from exponentiating the Faddeev-Popov
determinant. Since the FQFT algebra fails to close off-shell, we
need to introduce higher ghost terms into both the action and the
BRST transformation. In Feynman gauge, the local BRST Lagrangian
is
\begin{equation}
{\cal L}_{YM\,BRST}=-\frac{1}{2}\partial_\mu
A_{i\nu}\partial^\mu A^\nu_i
-\partial^\mu\bar{\eta_i}\partial_\mu\eta_i+ef_{ikl}\partial^\mu\bar{\eta}_i
A_{k\mu}\eta_l
$$ $$
+ef_{ikl}\partial_\mu
A_{i\nu}A^\mu_kA^\nu_l-\frac{1}{4}e^2f_{ikl}f_{lmn}
A_{i\mu}A_{k\nu}A^\mu_mA^\nu_n.
\end{equation}
It is invariant under the global symmetry transformation:
\[
\delta
A_{i\mu}=(\partial_\mu\eta_i-ef_{ikl}A_{k\mu}\eta_l)\delta\zeta,
\] \[
\delta\eta_i=-\frac{1}{2}ef_{ikl}\eta_k\eta_l\delta\zeta,
\]
\[
\delta\bar{\eta}_i=-\partial_\mu A^\mu_i\delta\zeta,
\]
where $\zeta$ is a constant anticommuting c-number.

The gluon and ghost kinetic operators are
\begin{equation}
{\cal K}^{\mu\nu}_{ik}=\delta_{ik}\eta^{\mu\nu}\partial^2,
{\cal K}_{ik}=\delta_{ik}\partial^2,
\end{equation}
The gluon propagator and the shadow gluon propagator are given by
\begin{equation}
D^{\mu\nu}_{ik}(p^2)=\frac{-i\delta_{ik}\eta^{\mu\nu}}{p^2-i\epsilon}
\exp\biggl(-p^2/\Lambda^2_{\rm YM}\biggr),
\end{equation}
\begin{equation}
D^{\rm shad\,\mu\nu}_{ik}(p^2)=\frac{-i\delta_{ik}\eta^{\mu\nu}}
{p^2-i\epsilon}\biggl[1-\exp\biggl(-p^2/\Lambda^2_{\rm YM}\biggr)
\biggr],
\end{equation}
where $\Lambda_{YM}$ denotes the FQFT Yang-Mills energy scale.

The regularized BRST action is
\begin{equation}
\hat{W}_{YM}[A,\bar{\eta},\eta]=\int d^4x\biggl\{-\frac{1}{2}
\partial_\nu\hat{A}_{i\mu}\partial^\nu\hat{A}_i^\mu-\frac{1}{2}B_{i\mu}
\bar{\cal O}^{-1}B^\mu_i
$$ $$
-\partial^\mu\hat{\bar{\eta}}_i\partial_\mu\hat{\eta}_i
-\bar{\chi}_i\bar{\cal O}^{-1}\chi_i\biggr\}+W^I_{\rm
YM}[A+B,\bar{\eta} +\bar{\chi},\eta+\chi],
\end{equation}
where $\chi$ is the ghost shadow field.

The regularizing, nonlocal BRST symmetry transformation is
\begin{equation}
\hat\delta A_{i\mu}=\bar{\cal E}^2\biggl\{(\partial_\mu\eta_i
+\partial_\mu\chi_i)-ef_{ikl}(A_{k\mu}+B_{k\mu})(\eta_l
+\chi_l)\biggr\}\delta\zeta,
$$ $$
\hat\delta\eta_i=-\frac{1}{2}ef_{ikl}\bar{\cal
E}^2(\eta_k+\chi_k) (\eta_l+\chi_l)\delta\zeta,
$$ $$
\hat{\delta}\bar{\eta}_i=-\bar{\cal E}^2(\partial_\mu A^\mu_i
+\partial_\mu B^\mu_i)\delta\zeta.
\end{equation}
The full functional, gauge fixed quantization is now given by
\begin{equation}
\langle 0\vert T^*(O[A,\bar{\eta},\eta])\vert 0\rangle_{\cal E}=\int[DA]
[D\bar{\eta}][D\eta]
\mu[A,\bar{\eta},\eta]O[\hat{A},\hat{\bar{\eta}},\hat{\eta}]
$$
$$
\times\exp(i\hat{W}_{\rm YM}[A,\bar{\eta},\eta]).
\end{equation}

Kleppe and Woodard~\cite{Kleppe2} have obtained the
invariant measure factor for the regularized Yang-Mills sector to
first order in the coupling constant $e$:
\begin{equation}
\ln(\mu[A,\bar{\eta},\eta])=-\frac{1}{2}e^2f_{ilm}f_{klm}\int
d^4x A_{i\mu}{\cal M}A^\mu_k+O(e^3),
\end{equation}
where
\begin{equation}
{\cal M}=\frac{1}{16\pi^2}\int^1_0d\tau
\frac{\Lambda^2}{(\tau+1)^2}\exp\biggl(\frac{\tau}{\tau+1}
\frac{\partial^2}{\Lambda^2}\biggr)
\biggl\{\frac{2+6\tau}{\tau+1}-3\biggr\}.
\end{equation}
The existence of a suitable invariant measure
factor implies that the necessary Slavnov-Taylor identities also
exist.

\section{Finite Perturbative Quantum Gravity}

As is well know, the problem with perturbative quantum gravity
based on a point-like graviton and a local field theory
formalism is that the theory is not
renormalizable~\cite{Veltman,Van}. Due to the Gauss-Bonnet
theorem, it can be shown that the one-loop graviton calculation
is renormalizable but two-loop is not~\cite{Sagnotti}.
Moreover, gravity-matter interactions are not renormalizable at
any loop order.

We shall now formulate the gravitational sector
in more detail as a FQFT. This problem has been considered
previously in the context of four-dimensional
GR~\cite{Moffat,Moffat2,Moffat4}. We shall expand the
gravity sector about flat Minkowski spacetime. In fact, FQFT can
be formulated as a perturbative theory by expanding around any
fixed, classical metric background~\cite{Veltman}
\begin{equation}
\label{background}
g_{\mu\nu}={\bar g}_{\mu\nu}+h_{\mu\nu},
\end{equation}
where ${\bar g}_{\mu\nu}$ is any smooth background metric field,
e.g. a de Sitter spacetime metric. For the sake of simplicity, we
shall only consider expansions about flat spacetime. Since the
gravitational field is weak up to the Planck energy scale, this
expansion is considered justified; even at the standard model
energy scale $E_{\rm SM}\sim 10^2$ GeV, we have $\kappa^2E^2_{\rm
SM}\sim 10^{-33}$.  Also, at these energy scales the curvature
of spacetime is very small. However, if we wish to include the
cosmological constant $\lambda$, then we cannot strictly speaking
expand about flat spacetime, because such an expansion of the
Einstein field equations will lead to the result that
$\lambda=0$. This is to be expected, because the cosmological
constant produces a curved spacetime even when the
energy-momentum tensor $T_{\mu\nu}=0$. Therefore, we should in
this case use the expansion (\ref{background}). But for energy
scales encountered in particle physics, the curvature is very
small, so we can approximate the perturbation caculation by using
the flat spacetime expansion and trust that the results are valid
in general for curved spacetime backgrounds including the
cosmological constant.

As in ref. [10], we will
regularize the GR equations using the covariant shadow field
formalism. Let us define ${\bf g}^{\mu\nu}=\sqrt{-g}g^{\mu\nu}$.
It can be shown that $\sqrt{-g}=\sqrt{-{\bf g}}$, where ${\bf g}=
{\rm det}({\bf g}^{\mu\nu})$ and $\partial_\rho{\bf g}={\bf
g}_{\alpha\beta}\partial_\rho{\bf g}^{\alpha\beta}{\bf g}$. We
can then write the local gravitational action $W_{\rm grav}$ in
the form~\cite{Goldberg}:
\begin{equation}
\label{action}
W_{\rm grav}=\int d^4x{\cal L}_{\rm grav}=\frac{1}{2\kappa^2}\int
d^4x [({\bf g}^{\rho\sigma}{\bf g}_{\lambda\mu} {\bf
g}_{\kappa\nu}
$$ $$
-\frac{1}{2}{\bf g}^{\rho\sigma} {\bf
g}_{\mu\kappa}{\bf g}_{\lambda\nu}
-2\delta^\sigma_\kappa\delta^\rho_\lambda{\bf
g}_{\mu\nu})\partial_\rho{\bf g}^{\mu\kappa} \partial_\sigma{\bf
g}^{\lambda\nu}
$$ $$
-\frac{1}{\alpha\kappa^2}\partial_\mu{\bf
g}^{\mu\nu}\partial_\kappa{\bf g}^{\kappa\lambda}
\eta_{\nu\lambda}
+{\bar C}^\nu\partial^\mu X_{\mu\nu\lambda}C^\lambda],
\end{equation}
where we have added a gauge fixing term with the parameter
$\alpha$, $C^\mu$ is the Fadeev-Popov ghost field and
$X_{\mu\nu\lambda}$ is a differential operator.

We expand the local interpolating graviton field ${\bf
g}^{\mu\nu}$ as
\begin{equation}
{\bf g}^{\mu\nu}=\eta^{\mu\nu}+\kappa\gamma^{\mu\nu}+O(\kappa^2).
\end{equation} Then,
\begin{equation}
{\bf g}_{\mu\nu}=\eta_{\mu\nu}-\kappa\gamma_{\mu\nu}
+\kappa^2{\gamma_\mu}^\alpha{\gamma_\alpha}_\nu+O(\kappa^3).
\end{equation}

The gravitational Lagrangian density is expanded as
\begin{equation}
{\cal L}_{\rm grav}={\cal L}^{(0)}+\kappa{\cal L}^{(1)}
+\kappa^2{\cal L}^{(2)}+....
\end{equation}
We obtain
\begin{equation}
{\cal L}^{(0)}=\frac{1}{2}\partial_\sigma\gamma_{\lambda\rho}
\partial^\sigma\gamma^{\lambda\rho}
-\partial_\lambda\gamma^{\rho\kappa}
\partial_\kappa\gamma^\lambda_\rho
-\frac{1}{4}\partial_\rho\partial^\rho\gamma
$$ $$
-\frac{1}{\alpha}\partial_\rho\gamma^\rho_\lambda\partial_\kappa
\gamma^{\kappa\lambda}
+{\bar C}^\lambda\partial_\sigma\partial^\sigma C_\lambda,
\end{equation}
\begin{equation}
{\cal L}^{(1)}
=\frac{1}{4}(-4\gamma_{\lambda\mu}\partial^\rho\gamma^{\mu\kappa}
\partial_\rho\gamma^\lambda_\kappa+2\gamma_{\mu\kappa}
\partial^\rho\gamma^{\mu\kappa}\partial_\rho\gamma
$$ $$
+2\gamma^{\rho\sigma}\partial_\rho\gamma_{\lambda\nu}
\partial_\sigma\gamma^{\lambda\nu}
-\gamma^{\rho\sigma}\partial_\rho\gamma\partial_\sigma\gamma
+4\gamma_{\mu\nu}\partial_\lambda\gamma^{\mu\kappa}
\partial_\kappa\gamma^{\nu\lambda})
$$ $$
+{\bar C}^\nu\gamma_{\kappa\mu}\partial^\kappa\partial^\mu C_\nu
+{\bar C}^\nu\partial^\mu\gamma_{\kappa\mu}\partial^\kappa C_\nu
-{\bar C}^\nu\partial^\lambda\partial^\mu\gamma_{\mu\nu}C_\lambda
-{\bar C}^\nu\partial^\mu\gamma_{\mu\nu}\partial^\lambda
C_\lambda,
\end{equation}
\begin{equation}
{\cal L}^{(2)}=\frac{1}{4}(4\gamma_{\kappa\alpha}
\gamma^{\alpha\nu}
\partial^\rho\gamma^{\lambda\kappa}\partial_\rho\gamma_{\nu\lambda}
+(2\gamma_{\lambda\mu}\gamma_{\kappa\nu}-\gamma_{\mu\kappa}\gamma_{\nu\lambda})
\partial^\rho\gamma^{\mu\kappa}\partial_\rho\gamma^{\nu\lambda}
$$ $$
-2\gamma_{\lambda\alpha}\gamma^\alpha_\nu\partial^\rho\gamma^{\lambda\nu}
\partial_\rho\gamma-2\gamma^{\rho\sigma}\gamma^\kappa_\nu
\partial_\rho\gamma_{\lambda\kappa}\partial_\sigma\gamma^{\nu\lambda}
$$ $$
+\gamma^{\rho\sigma}\gamma^{\nu\lambda}\partial_\sigma\gamma_{\nu\lambda}
\partial_\rho\gamma-2\gamma_{\mu\alpha}\gamma^{\alpha\nu}
\partial^\lambda\gamma^{\mu\kappa}\partial_\kappa\gamma_{\nu\lambda}),
\end{equation}
where $\gamma={\gamma^\alpha}_\alpha$.

In the limit $\alpha\rightarrow\infty$, the Lagrangian density
${\cal L}_{\rm grav}$ is invariant under the gauge
transformation
\begin{equation}
\delta\gamma_{\mu\nu}=X_{\mu\nu\lambda}\xi^\lambda,
\end{equation}
where $\xi^\lambda$ is an infinitesimal vector quantity and
\begin{equation}
X_{\mu\nu\lambda}=\kappa(-\partial_\lambda\gamma_{\mu\nu}
+2\eta_{(\mu\lambda}\gamma_{\kappa\nu)}\partial^\kappa)
+(\eta_{(\mu\lambda}\partial_{\nu)}-\eta_{\mu\nu}\partial_\lambda).
\end{equation}
However, for the quantized theory it is more useful to require
the BRST symmetry. We choose $\xi^\lambda=C^\lambda\sigma$,
where $\sigma$ is a global anticommuting scalar. Then, the BRST
transformation is
\begin{equation}
\delta\gamma_{\mu\nu}=X_{\mu\nu\lambda}C^\lambda\sigma,
\quad \delta {\bar C}^\nu=-\partial_\mu\gamma^{\mu\nu}
\biggl(\frac{2\sigma}{\alpha}\biggr),\quad \delta C_\nu=\kappa
C^\mu \partial_\mu C_\nu\sigma.
\end{equation}

We now substitute the operators
\begin{equation}
\gamma_{\mu\nu}\rightarrow{\hat\gamma}_{\mu\nu},\quad
C_\lambda\rightarrow {\hat C}_\lambda,\quad
{\bar C}_\nu\rightarrow {\hat{\bar C}}_\nu,
\end{equation}
where
\begin{equation}
\label{hatgamma}
{\hat \gamma}_{\mu\nu}={\cal E}^{-1}\gamma_{\mu\nu},
\quad {\hat C}_\lambda={\cal E}^{-1}C_\lambda,\quad
{\hat{\bar C}}_\lambda={\cal E}^{-1}C_\lambda.
\end{equation}

As in the case of the Yang-Mills sector, the on shell propagators
are unaltered from their local antecedents, while virtual
particles are nonlocal. This destroys the gauge invariance
of e.g. graviton-graviton scattering and requires an
iteratively defined series of ``stripping'' vertices to
ensure the decoupling of all unphysical modes. Moreover, the
local gauge transformations have to be extended to nonlinear,
nonlocal gauge transformations to guarantee the over-all
invariance of the regularized amplitudes. Cornish has derived
the primary graviton vertices and the BRST symmetry relations for
the regularized ${\hat W}_{\rm grav}$~\cite{Cornish,Cornish2},
using the nonlinear, nonlocal extended gauge transformations
suitable for the perturbative gravity equations.

The regularized graviton propagator in the fixed
de Donder gauge $\alpha=-1$~\cite{Donder} is given by
\begin{equation}
D^{\rm grav}_{\mu\nu\rho\sigma}(x)
=(\eta_{\mu\rho}\eta_{\nu\sigma}+\eta_{\mu\sigma}\eta_{\nu\rho}
-\eta_{\mu\nu}\eta_{\rho\sigma})
$$ $$
\times\biggl(\frac{-i}{(2\pi)^4}\biggr)\int d^4k\frac{{\cal
E}^2(k^2)}{k^2-i\epsilon} \exp[ik\cdot(x-x')],
\end{equation}
while the shadow propagator is
\begin{equation}
D^{\rm shad}_{\mu\nu\rho\sigma}(x)
=(\eta_{\mu\rho}\eta_{\nu\sigma}+\eta_{\mu\sigma}\eta_{\nu\rho}
-\eta_{\mu\nu}\eta_{\rho\sigma})
$$ $$
\times\biggl(\frac{-i}{(2\pi)^4}\biggr)\int d^4k \frac{[1-{\cal
E}^2(k^2)]}{k^2-i\epsilon} \exp[ik\cdot(x-x')].
\end{equation}
The ghost propagator in momentum space is given by
\begin{equation}
D^G_{\mu\nu}(p)=\frac{\eta_{\mu\nu}{\cal E}^2(p^2)}{p^2},
\end{equation}
while the shadow ghost propagator is
\begin{equation}
D^{\rm shad\,G}_{\mu\nu}(p)=\frac{\eta_{\mu\nu}
[1-{\cal E}^2(p^2)]}{p^2}.
\end{equation}

In momentum space we have
\begin{equation}
\frac{-i{\cal E}^2(k^2)}{k^2-i\epsilon}
=-i\int^{\infty}_1\frac{d\tau}{\Lambda^2_G}
\exp\biggl(-\tau\frac{k^2}{\Lambda^2_G}\biggr),
\end{equation}
and
\begin{equation}
\frac{i({\cal E}^2(k^2)-1)}{k^2-i\epsilon}=-i\int_0^1\frac{d\tau}
{\Lambda^2_G}
\exp\biggl(-\tau\frac{k^2}{\Lambda^2_G}\biggr),
\end{equation}
where $\Lambda_G$ is the gravitational scale parameter.
 
The local propagator is reproduced by subtracting $D^{\rm
shad}$ from $D^{\rm grav}$, while the ``stripped'' vertices
are obtained by subtracting the amplitudes containing the shadow
propagator $D^{\rm shad\,}$ from the amplitudes containing the
regulator operators (\ref{hatgamma}).
We can facilitate the calculations by separating the free and
interacting parts of the action
\begin{equation}
W_{\rm grav}(\gamma)=W^F_{\rm grav}(\gamma)
+W^I_{\rm grav}(\gamma).
\end{equation}
The finite regularized gravitational action is given by
\begin{equation}
\hat{W}_{\rm grav}(\gamma,s)=W^F_{\rm grav}({\hat \gamma})
-P_{\rm grav}(s)+W^I_{\rm grav}(\gamma+s),
\end{equation}
where
\begin{equation}
{\hat \gamma}={\cal E}^{-1}\gamma,\quad P_{\rm grav}(s)
=\int d^4x{\cal G}(\sqrt{s},s_i{\cal O}_{ij}^{-1}s_j),
\end{equation}
$s$ denotes the graviton shadow field, and ${\cal G}$ denotes
the detailed expansion of the contributions formed from the
shadow field.
 
The regularized
Lagrangian density up to order $\kappa^2$ is invariant under the
extended BRST transformations~\cite{Cornish}:
\begin{equation}
{\hat\delta}_0\gamma_{\mu\nu}=X^{(0)}_{\mu\nu\lambda}
C^\lambda\sigma=(\partial_\nu C_\mu+\partial_\mu
C_\nu-\eta_{\mu\nu}\partial_\lambda C^\lambda)\sigma,
\end{equation}
\begin{equation}
{\hat\delta}_1\gamma_{\mu\nu}=\kappa{\cal
E}^2X^{(1)}_{\mu\nu\lambda} C^\lambda\sigma
=\kappa{\cal E}^2(2\gamma_{\rho(\mu}\partial^\rho C_{\nu)}
-\partial_\lambda\gamma_{\mu\nu}
C^\lambda-\gamma_{\mu\nu}\partial_\lambda C^\lambda),
\end{equation}
\begin{equation}
{\hat\delta}_0{\bar C}^\nu=2\partial_\mu\gamma^{\mu\nu}\sigma,
\end{equation}
\begin{equation}
{\hat\delta}_1C_\nu=\kappa{\cal E}^2C^\mu \partial_\mu
C_\nu\sigma.
\end{equation}

The order $\kappa^2$ transformations are
\begin{equation}
{\hat\delta}_2\gamma_{\mu\nu}=\kappa^2{\cal E}^2[2\partial^\rho
C_{(\nu}
D^{\rm shad}_{\mu)\rho\kappa\lambda}
(B^{\kappa\lambda}+H^{\kappa\lambda})
$$ $$
-C^\rho D^{\rm shad}_{\mu\nu\kappa\lambda}(\partial_\rho
B^{\kappa\lambda} +\partial_\rho H^{\kappa\lambda})
-\partial_\rho C^\rho D^{\rm shad}_{\mu\nu\kappa\lambda}
(B^{\kappa\lambda}+H^{\kappa\lambda})
$$ $$
+2\gamma_{\rho(\mu}
D^{\rm shad\,G}_{\nu)\kappa}\partial^\rho H^\kappa
-\partial_\rho\gamma_{\mu\nu}D^{\rm shad\,G\,\rho\kappa}H_\kappa
-\gamma_{\mu\nu}D^{\rm shad\,G\,\rho\kappa}\partial_\rho
H_\kappa]\sigma,
\end{equation}
\begin{equation}
{\hat\delta}_2C_\nu=-\kappa^2{\cal E}^2(\partial_\mu C_\nu
D^{\rm shad\,G\,\rho\kappa}H_\kappa+C_\mu
D^{\rm shad\,G}_{\nu\kappa}\partial^\mu H^\kappa)\sigma.
\end{equation}
Here, we have
\begin{equation}
H^{\alpha\beta}=-(\partial^{(\alpha}{\bar C}_\rho
\partial^{\beta)}C^\rho+\partial^\rho{\bar C}^{(\alpha}
\partial^{\beta)}C_\rho
+\partial^\rho\partial^{(\beta}{\bar C}^{\alpha)} C_\rho),
\end{equation}
\begin{equation}
H^\rho=\gamma_{\lambda\kappa}\partial^\lambda\partial^\kappa
C^\rho+\partial^\kappa\gamma_{\lambda\kappa}\partial^\lambda
C^\rho
-\partial_\kappa\partial_\lambda\gamma^{\rho\kappa}
C^\lambda- \partial_\kappa\gamma^{\rho\kappa}\partial_\lambda
C^\lambda,
\end{equation}
\begin{equation}
{\bar H}^\rho
=\partial^\lambda{\bar C}^\rho\partial^
\kappa\gamma_{\lambda\kappa}
+\partial^\lambda\partial^\kappa{\bar
C}^\rho\gamma_{\lambda\kappa} +\partial^\rho{\bar
C}^\lambda\partial^\kappa \gamma_{\lambda\kappa}.
\end{equation}

Because we have extended the gauge symmetry to nonlinear,
nonlocal transformations, we must also supplement the
quantization procedure with an invariant measure
\begin{equation}
{\cal M}=\Delta({\bf g}, {\bar C}, C)D[{\bf
g}_{\mu\nu}]D[{\bar C}_\lambda]D[C_\sigma]
\end{equation}
such that $\delta {\cal M}=0$.

As we have demonstrated, the quantum gravity perturbation theory
is invariant under the FQFT generalized, nonlinear field
representation dependent transformations. It is unitary and
finite to all orders in a way similar to the non-Abelian gauge
theories formulated using FQFT. At the tree graph level all
unphysical polarization states are decoupled and nonlocal effects
will only occur in graviton and graviton-matter loop graphs.
Because the gravitational tree graphs are purely local there is a
well-defined classical GR limit. The finite quantum gravity
theory is well-defined in four real spacetime dimensions.

We quantize by means of the path integral operation
\begin{equation}
\langle 0\vert T^*(O[{\bf g}])\vert 0\rangle_{\cal E}=\int[D{\bf g}]
\mu[{\bf g}]({\rm gauge\, fixing})
O[\hat {\bf g}]\exp(i\hat W_{\rm grav}[{\bf g}]).
\end{equation}
The quantization is carried out
in the functional formalism by finding a measure factor
$\mu[{\bf g}]$ to make $[D{\bf g}]$ invariant under the
classical symmetry. To ensure a correct gauge fixing scheme, we write
$W_{\rm grav}[{\bf g}]$ in the BRST invariant form with ghost fields; the
ghost structure arises from exponentiating the Faddeev-Popov
determinant~\cite{Fradkin}.
The algebra of extended gauge symmetries is not expected to close
off-shell, so one needs to introduce higher ghost terms (beyond
the normal ones) into both the action and the BRST
transformation. The BRST action will be regularized directly to
ensure that all the corrections to the measure factor are
included.

\section{\bf A Resolution of The Higgs Hierarchy Problem}

It is time to discuss the Higgs sector hierarchy
problem~\cite{Susskind}. The gauge hierarchy problem is
related to the spin $0^+$ scalar field nature of the Higgs
particle in the standard model with quadratic mass divergence and
no protective extra symmetry at $m=0$. In standard point
particle, local field theory the fermion masses are
logarithmically divergent and there exists a chiral symmetry
restoration at $m=0$.
Writing $m_H^2=m_{0H}^2+\delta m_H^2$, where
$m_{0H}$ is the bare Higgs mass and $\delta m_H$ is the Higgs
self-energy renormalization constant, we get for the one loop Feynman
graph in $D=4$ spacetime:
\begin{equation}
\delta m_H^2\sim \frac{g}{32\pi^2}M_c^2,
\end{equation}
where $M_c$ is a cutoff parameter. If we want to understand
the nature of the Higgs mass we must require that
\begin{equation}
\delta m_H^2 \leq O(m_H^2),
\end{equation}
i.e. the quadratic divergence should be cut off at the mass scale of
the order of the physical Higgs mass. Since $m_H\simeq \sqrt{2g}v$,
where $v=<\phi>_0$ is the vacuum expectation value of the scalar field
$\phi$ and $v=246$ GeV from the electroweak theory, then in order to keep
perturbation theory valid, we must demand that $10\,{\rm GeV}
\leq m_H \leq 350\,{\rm GeV}$ and we need
\begin{equation}
\label{Higgscut}
M_c =M_{\rm Higgs}\leq 1\, {\rm TeV},
\end{equation}
where the lower bound on $m_H$ comes from the avoidance of washing out
the spontaneous symmetry breaking of the vacuum.

Nothing in the standard model can tell us why (\ref{Higgscut}) should be true,
so we must go beyond the local standard model to solve the problem.
$M_c$ is an arbitrary parameter in point particle field theory
with no physical interpretation. Since all particles interact
through gravity, then ultimately we should expect to include
gravity in the standard model, so we expect that $M_{\rm
Planck}\sim 10^{19}$ GeV should be the natural cutoff. Then we
have using (\ref{Higgscut}) and $g\sim 1$:
\[
\frac{\delta m_H^2(M_{\rm Higgs})}{\delta m_H^2
(M_{\rm Planck})} \approx \frac{M^2_{\rm Higgs}}
{M^2_{\rm Planck}}\approx 10^{-34}, \]
which represents an intolerable fine-tuning of parameters. This `naturalness' or
hierarchy problem is one of the most serious defects of the standard model.

There have been two strategies proposed as ways out of the hierarchy
problem. The Higgs is taken to be composite at a scale
$M_c\simeq 1$ TeV, thereby providing a natural cutoff in
the quadratically divergent Higgs loops. One such scenario is the
`technicolor' model, but it cannot be reconciled with the
accurate standard model data, nor with the smallness of fermion
masses and the flavor-changing neutral current interactions. The
other strategy is to postulate supersymmetry, so that the
opposite signs of the boson and fermion lines cancel by means of
the non-renormalization theorem. However, supersymmetry is badly
broken at lower energies, so we require that
\[
\delta m_H^2\sim \frac{g}{32\pi^2}\vert M^2_{c\,{\rm bosons}}
-M^2_{c\,{\rm fermions}}\vert\leq 1\,{\rm TeV}^2,
\]
or, in effect
\[
\vert m_b-m_f\vert \leq 1\, {\rm TeV}.
\]
This physical requirement leads to the prediction that the supersymmetric
partners of known particles should have a threshold $\leq1$ TeV.

A third possible strategy is to introduce the FQFT formalism, and realize a
field theory mechanism which will introduce a natural physical scale in the
theory $\Lambda_H\leq 1$ TeV, which will protect the Higgs
mass from becoming large and unstable.

Let us consider the regularized scalar field FQFT Lagrangian in Minkowski spacetime
\begin{equation}
{\hat{\cal L}}_S=\frac{1}{2}\hat\phi(\partial^2-m^2)\hat\phi
-\frac{1}{2}\rho{\cal O}^{-1}\rho+\frac{1}{2}Z^{-1}\delta m^2(\phi+\rho)^2
-\frac{1}{24}g_0(\phi+\rho)^4,
\end{equation}
where $\phi=Z^{1/2}\phi_R$ is the bare field, $\phi_R$ is the renormalized
field, $\hat\phi={\cal E}^{-1}\phi$, $\rho$ is
the shadow field, $m_0$ is the bare mass, $Z$ is the field strength
renormalization constant, $\delta m^2$ is the mass renormalization constant
and $m$ is the physical mass. The regularizing operator is given by
\begin{equation}
{\cal E}_m=\exp\biggl(\frac{\partial^2-m^2}{2\Lambda_H^2}\biggr),
\end{equation}
while the shadow kinetic operator is
\begin{equation}
{\cal O}^{-1}=\frac{\partial^2-m^2}{{\cal E}_m^2-1}.
\end{equation}
Here, $\Lambda_H$ is the Higgs scalar field energy scale in FQFT,
which determines the scale of nonlocalizability of the Higgs
particle.

The full propagator is
\begin{equation}
-i\Delta_R(p^2)=\frac{-i{\cal E}_m^2}{p^2+m^2-i\epsilon}
=-i\int_1^{\infty}\frac{d\tau}{\Lambda_H^2}\exp\biggl[-\tau
\biggl(\frac{p^2+m^2}{\Lambda_H^2}\biggr)\biggr],
\end{equation}
whereas the shadow propagator is
\begin{equation}
i\Delta_{\rm shadow}
=i\frac{{\cal E}_m^2-1}{p^2+m^2}=-i\int^1_0\frac{d\tau}{\Lambda_H^2}
\exp\biggl[-\tau\biggl(\frac{p^2+m^2}{\Lambda_H^2}\biggr)\biggr].
\end{equation}

Let us define the self-energy $\Sigma(p)$ as a Taylor series expansion
around the mass shell
$p^2=-m^2$:
\begin{equation}
\Sigma(p^2)=\Sigma(-m^2)+(p^2+m^2)\frac{\partial\Sigma}{\partial p^2}(-m^2)
+{\tilde \Sigma}(p^2),
\end{equation}
where ${\tilde\Sigma}(p^2)$ is the usual finite part in the point
particle limit $\Lambda_H\rightarrow\infty$.
We have
\begin{equation}
{\tilde\Sigma}(-m^2)=0,
\end{equation}
and
\begin{equation}
\frac{\partial{\tilde\Sigma}(p^2)}{\partial p^2}(p^2=-m^2)=0.
\end{equation}

The full propagator is related to the self-energy $\Sigma(p^2)$ by
\begin{equation}
-i\Delta_R(p^2)=\frac{-i{\cal E}_m^2[1+{\cal O}\Sigma(p^2)]}{p^2+m^2
+\Sigma(p^2)}
=\frac{-iZ}{p^2+m^2+\Sigma_R(p^2)}.
\end{equation}
Here $\Sigma_R(p^2)$ is the renormalized self-energy which can be written
as
\begin{equation}
\Sigma_R(p^2)=(p^2+m^2)\biggl[\frac{Z}{{\cal E}_m^2(1+{\cal O}\Sigma)}
-1\biggr]+\frac{Z\Sigma}{{\cal E}_m^2(1+{\cal O}\Sigma)}.
\end{equation}
The 1PI two-point function is given by
\begin{equation}
-i\Gamma_R^{(2)}(p^2)=i[\Delta_R(p^2)]^{-1}=\frac{i[p^2+m^2+\Sigma(p^2)]}
{{\cal E}_m^2[1+{\cal O}\Sigma(p^2)]}.
\end{equation}
Since ${\cal E}_m\rightarrow 1$ and ${\cal O}\rightarrow 0$ as $\Lambda_H
\rightarrow\infty$,
then in this limit
\begin{equation}
-i\Gamma_R^{(2)}(p^2)=i[p^2+m^2+\Sigma(p^2)],
\end{equation}
which is the standard point particle result.

The mass renormalization is determined by the propagator pole at $p^2=-m^2$
and we have
\begin{equation}
\Sigma_R(-m^2)=0.
\end{equation}
Also, we have the condition
\begin{equation}
\frac{\partial\Sigma_R(p^2)}{\partial p^2}(p^2=-m^2)=0.
\end{equation}

The renormalized coupling constant is defined by the four-point function
$\Gamma^{(4)}_R(p_1,p_2,p_3,p_4)$ at the point $p_i=0$:
\begin{equation}
\Gamma_R^{(4)}(0,0,0,0)=g.
\end{equation}
The bare coupling constant $g_0$ is determined by
\begin{equation}
Z^2g_0=g+\delta g(g,m^2,\Lambda_H^2).
\end{equation}
Moreover,
\[
Z=1+\delta Z(g,m^2,\Lambda_H^2),
\]
\[
Zm_0^2=Zm^2-\delta m^2(g,m^2,\Lambda^2_H).
\]

A calculation of the scalar field mass renormalization in
D-dimensional space gives
\cite{Woodard2}:
\begin{equation}
\delta m^2=\frac{g}{2^{D+1}\pi^{D/2}}m^{D-2}
\Gamma\biggl(1-\frac{D}{2},\frac{m^2}{\Lambda_H^2}\biggr)+O(g^2),
\end{equation}
where $\Gamma(n,z)$ is the incomplete gamma function:
\begin{equation}
\Gamma(n,z)=\int_z^{\infty}\frac{dt}{t}t^n\exp(-t)=(n-1)\Gamma(n-1,z)
+z^{n-1}\exp(-z).
\end{equation}

We have
\begin{equation}
\Gamma(-1,z)=-E_i(z)+\frac{1}{z}\exp(-z),
\end{equation}
where $E_i(z)$ is the exponential integral
\[
E_i(z)\equiv \int^\infty_zdt\frac{\exp(-t)}{t}.
\]
For small $z$ we obtain the expansion
\begin{equation}
E_i(z)=-\ln(z)-\gamma+z-\frac{z^2}{2\cdot 2!}+\frac{z^3}{3\cdot 3!}-...,
\end{equation}
where $\gamma$ is Euler's constant. For large positive values of $z$, we
have the asymptotic expansion
\begin{equation}
E_i(z)\sim\exp(-z)\biggl[\frac{1}{z}-\frac{1}{z^2}+\frac{2!}{z^3}-...\biggr].
\end{equation}
Thus, for small $m/\Lambda_H$ we obtain in $D=4$ spacetime:
\begin{equation}
\label{lambdast}
\delta m^2=\frac{g}{32\pi^2}
\biggl[\Lambda_H^2-m^2\ln\biggl(\frac{\Lambda_H^2}{m^2}\biggr)
-m^2(1-\gamma)+O\biggl(\frac{m^2}{\Lambda_H^2}\biggr)\biggr]+O(g^2),
\end{equation}
which is the standard quadratically divergent self-energy, obtained from
a cutoff procedure or a dimensional regularization scheme.

We have for $z\rightarrow\infty$:
\begin{equation}
\label{Gamma}
\Gamma(a,z)\sim
z^{a-1}\exp(-z)\biggl[1+\frac{a-1}{z}
+O\biggl(\frac{1}{z^2}\biggr)\biggr]
\end{equation}
so that for $m\gg\Lambda_H$, we get in four-dimensional spacetime
\begin{equation}
\delta
m^2\sim\frac{g}{32\pi^2}\biggl(\frac{\Lambda^4_H}{m^2}\biggr)
\exp\biggl(-\frac{m^2}{\Lambda_H^2}\biggr).
\end{equation}
Thus, the Higgs self-energy one loop graph falls off
exponentially fast for $m\gg \Lambda_H$. We have succeeded
in stabilizing the radiative corrections to the Higgs sector,
solving the Higgs hierarchy problem for $\Lambda_H\leq 1$ TeV.

\section{Gluon and Gravitational Vacuum Polarization}

A calculation of the one-loop gluon vacuum polarization in FQFT
gives the tensor in D-dimensions
\begin{equation}
\Pi^{\mu\nu}_{ik}(p)=\frac{g^2}{2^Dp^{D/2}}f_{ilm}f_{klm}
(p^2\eta^{\mu\nu}-p^\mu p^\nu)\Pi(p^2),
\end{equation}
where $p$ is the gluon momentum and
\begin{equation}
\Pi(p^2)=2\exp\biggl(-p^2/\Lambda_{\rm YM}^2\biggr)
\int^{1/2}_0dy\Gamma(2-D/2, yp^2/\Lambda_{\rm YM}^2)
[y(1-y)p^2]^{D/2-2}
$$ $$
\times[2(D-2)y(1-y)-\frac{1}{2}(D-6)].
\end{equation}
We observe that $\Pi^\mu_{ik\,\mu}(0)=0$ a result that is
required by gauge invariance and the fact that the gluon has zero
mass.

The dimensionally regulated gluon vacuum polarization result is
obtained by the replacement
\begin{equation}
\Gamma(2-D/2,yp^2/\Lambda^2_{\rm YM})\rightarrow \Gamma(2-D/2)
\end{equation}
and choosing $p^2 \ll \Lambda^2_{\rm YM}$. In four-dimensions we
get
\begin{equation}
\label{vacpolarization}
\Pi(p^2)=2\exp(-p^2/\Lambda^2_{\rm YM})\int^{1/2}_0
dyE_i(yp^2/\Lambda^2_{\rm YM})[4y(1-y)+1],
\end{equation}
where we have used the relation
\begin{equation}
\Gamma(0,z)\equiv E_i(z)=\int^\infty_zdt\exp(-t)t^{-1}.
\end{equation}
By using the behavior for large Euclidean momentum $p^2 \gg
\Lambda^2_{\rm YM}$:
\begin{equation}
E_i(yp^2/\Lambda^2_{\rm YM})\sim
\frac{\Lambda^2_{\rm YM}}{p^2}
\exp(-yp^2/\Lambda^2_{\rm YM}),
\end{equation}
we find from (\ref{vacpolarization}) that
\begin{equation}
\Pi(p^2)\sim \frac{\Lambda^2_{\rm YM}}{p^2}
\exp\biggl(-p^2/\Lambda^2_{\rm YM}\biggr)
\biggl[\exp(-p^2/2\Lambda^2_{\rm YM})
+\frac{4\Lambda^2_{\rm YM}}{p^2}
$$ $$
+\frac{\Lambda^4_{\rm YM}}{p^4} -\frac{16\Lambda^6_{\rm
YM}}{p^6}\biggr].
\end{equation}
Thus, the gluon vacuum polarization is exponentially damped for
$p^2\gg\Lambda^2_{\rm YM}$.

The lowest order
contributions to the graviton self-energy in FQFT will include
the standard graviton loops, the shadow field graviton loops, the
ghost field loop contributions with their shadow field
counterparts, and the measure loop contributions. In the
regularized perturbative gravity theory the first order vacuum
polarization tensor $\Pi^{\mu\nu\rho\sigma}$ must satisfy the
Slavnov-Ward identities~\cite{Medrano}:
\begin{equation}
\label{SlavnovWard}
p_\mu
p_\rho
D^{\mu\nu\alpha\beta}(p)\Pi_{\alpha\beta\gamma\delta}(p)
D^{\gamma\delta\rho\sigma}(p)=0.
\end{equation}
By symmetry and Lorentz invariance, the vacuum polarization
tensor must have the form
\begin{equation}
\Pi_{\alpha\beta\gamma\delta}(p)
=\Pi_1(p^2)p^4\eta_{\alpha\beta}\eta_{\gamma\delta}+\Pi_2
(p^2)p^4(\eta_{\alpha\gamma}\eta_{\beta\delta}
+\eta_{\alpha\delta}\eta_{\beta\gamma})
$$ $$
+\Pi_3(p^2)p^2(\eta_{\alpha\beta}p_\gamma
p_\delta+\eta_{\gamma\delta}p_\alpha p_\beta)
+\Pi_4(p^2)p^2(\eta_{\alpha\gamma}p_\beta
p_\delta+\eta_{\alpha\delta}p_\beta p_\gamma
$$ $$
+\eta_{\beta\gamma}p_\alpha p_\delta+\eta_{\beta\delta}p_\alpha
p_\gamma)+\Pi_5(p^2)p_\alpha p_\beta p_\gamma p_\delta.
\end{equation}
The Slavnov-Ward identities impose the restrictions
\begin{equation}
\Pi_2+\Pi_4=0,\quad 4(\Pi_1+\Pi_2-\Pi_3)+\Pi_5=0.
\end{equation}

The basic lowest order graviton self-energy diagram is determined
by~\cite{Leibbrandt,Brown,Donoghue,Duff,Duff2}:
\begin{equation}
\Pi^1_{\mu\nu\rho\sigma}(p)=\frac{1}{2}\kappa^2\exp\biggl(-p^2/\Lambda^2_G\biggr)\int
d^4q {\cal U}_{\mu\nu\alpha\beta\gamma\delta}(p,-q,q-p)
D^{\alpha\beta\kappa\lambda}(q)
$$ $$
\times D^{\gamma\delta\tau\xi}(p-q){\cal
U}_{\kappa\lambda\tau\xi\rho\sigma}(q,p-q,-p),
\end{equation}
where ${\cal U}$ is the three-graviton vertex function
\begin{equation}
{\cal U}_{\mu\nu\rho\sigma\delta\tau}(q_1,q_2,q_3) =
-\frac{1}{2}[q_{2(\mu}q_{3\nu)}\biggl(2\eta_{\rho(\delta}\eta_{\tau)\sigma}
-\frac{2}{D-2}\eta_{\mu\nu}\eta_{\delta\tau}\biggr)
$$ $$
+q_{1(\rho}q_{3\sigma)}\biggl(2\eta_{\mu(\delta}\eta_{\tau)\nu}
-\frac{2}{D-2}\eta_{\mu\nu}\eta_{\delta\tau}\biggr)+...],
\end{equation}
and the ellipsis denote similar contributions.

To this diagram, we must add the ghost particle diagram
contribution $\Pi^2$, the shadow diagram contribution $\Pi^3$ and
the measure diagram contribution $\Pi^4$. The dominant finite
contribution to the graviton self-energy will be of the form
\begin{equation}
\label{gravpol}
\Pi_{\mu\nu\rho\sigma}(p)\sim\kappa^2\Lambda_G^4\exp\biggl(-p^2/\Lambda^2_G\biggr)
Q_{\mu\nu\rho\sigma}(p^2)
$$ $$
\sim\frac{\Lambda^4_G}
{M^2_{\rm PL}}\exp\biggl(-p^2/\Lambda^2_G\biggr)Q_{\mu\nu\rho\sigma}(p^2),
\end{equation}
where $M_{\rm PL}$ is the reduced Planck mass and $Q(p^2)$ is a
finite remaining part.

For {\it renormalizable} field theories such as quantum
electrodynamics and Yang-Mills theory, we will find that in FQFT
the loop contributions are controlled by the incomplete
$\Gamma$-function. If we adopt an ``effective'' quantum gravity
theory expansion in the energy~\cite{Donoghue}, then we would
expect to obtain
\begin{equation}
\Pi_{\mu\nu\rho\sigma}(p)\sim\kappa^2
\exp\biggl(-p^2/\Lambda_G^2\biggr)
{\cal F}(\Gamma(2-D/2,p^2/\Lambda_G^2)
Q_{\mu\nu\rho\sigma}(p^2),
\end{equation}
where ${\cal F}$ denotes the functional dependence on the
incomplete $\Gamma$-function. By making the replacement
\begin{equation}
{\cal F}(\Gamma(2-D/2),p^2/\Lambda^2_G)\rightarrow
{\cal F}(\Gamma(2-D/2)),
\end{equation}
we would then obtain the second order graviton loop calculations
using dimensional regularization
~\cite{Leibbrandt,Brown,Donoghue,Duff,Duff2,Veltman}.
The dominant behavior will now be $\ln(\Lambda^2_G/q^2)$ and
not $\Lambda^4_G$. However, in a nonrenormalizable theory
such as quantum gravity, the dimensional
regularization technique may not provide a correct result for the
dominant behavior of the loop integral and we expect the result
to be of order $\Lambda_G^4$. Indeed, it is well known that
dimensional regularization for massless particles removes all
contributions from tadpole graphs and $\delta^4(0)$ contact
terms. On the other hand, FQFT takes into account all leading
order contributions and provides a complete account of all
counterterms. Because all the scattering amplitudes are finite,
then renormalizability is no longer an issue.

The function
\begin{equation}
{{Q_{\mu}}^{\mu\sigma}}_\sigma(p^2)\sim p^4
\end{equation}
as $p^2\rightarrow 0$. Therefore, $\Pi_{\mu\nu\rho\sigma}(p)$
vanishes at $p^2=0$ as it should from gauge invariance and for
massless gravitons.

In Euclidean momentum space, which we can reach by a Wick
rotation, we see that for $p^2\gg \Lambda_G^2$ the graviton
self-energy (\ref{gravpol}) is exponentially damped and the
quantum gravity loop corrections are negligible for energies
greater than $\Lambda_G$.

It is often argued in the literature on quantum
gravity that the gravitational quantum corrections scale as
$\alpha_G =GE^2$, so that for sufficiently large values of the
energy $E$, namely, of order the Planck energy, the gravitational
quantum fluctuations become large. We see that in FQFT this will
not be the case, because the finite quantum loop corrections
become negligible in the high energy limit provided the
perturbative approximation is valid. Of course, the contributions
of the tree graph exchanges of virtual gravitons can be large in
the high energy limit, corresponding to strong classical
gravitational fields. It follows that for high enough energies, a
classical curved spacetime would be a good approximation, at
least until the perturbation calculations break down.

In contrast to recent models of branes and strings in which the
higher-dimensional compactification scale is lowered to the TeV
range~\cite{Witten}, we retain the classical GR
gravitation picture and its Newtonian limit. It is perhaps a
radical notion to entertain that quantum gravity becomes weaker
as the energy scale increases towards the Planck scale $\sim
10^{19}$ Gev, but there is, of course, no known experimental
reason why this should not be the case in nature. However, we do
not expect that our weak gravity field expansion is valid at the
Planck scale when $GE^2\sim 1$, although the exponential damping
of the quantum gravity loop graphs could still persist at the
Planck scale. This question remains unresolved until a
nonperturbative solution to quantum gravity is found.

It is worth noting that in the framework of an effective
gravitational field theory~\cite{Donoghue}, the leading
lowest order loop divergence can be ``renormalized'' by being
absorbed into two parameters $c_1$ and $c_2$. For a non-flat
spacetime background metric ${\bar g}_{\mu\nu}$, the divergent
term at one loop due to graviton and ghost loops is given
by~\cite{Veltman}: \begin{equation}
{\cal L}^{\rm div}_{1{\rm loop}}=\frac{1}{8\pi^2\epsilon}
\biggl[\frac{1}{120}{\bar R}^2+\frac{7}{20}{\bar R}_{\mu\nu},
{\bar R}^{\mu\nu}\biggr],
\end{equation}
where $\epsilon=4-D$ and the effective field theory
renormalization parameters are
\begin{equation}
c^{(r)}_1=c_1+\frac{1}{960\pi^2\epsilon},\quad
c^{(r)}_2=c_2+\frac{7}{160\pi^2\epsilon}.
\end{equation}
 
\section{A Quantum Gravity Resolution of the Cosmological Constant
Problem}

Zeldovich~\cite{Zeldovich} showed that the zero-point
vacuum fluctuations must have a Lorentz invariant form
\begin{equation}
T_{{\rm vac}\,\mu\nu}=\lambda_{\rm vac}g_{\mu\nu},
\end{equation}
consistent with the equation of state $\rho_{\rm vac}=-p_{\rm
vac}$. Thus, the vacuum within the framework of particle quantum
physics has properties identical to the cosmological constant.
In quantum theory, the second quantization of a classical field
of mass $m$, treated as an ensemble of oscillators each with a
frequency $\omega(k)$, leads to a zero-point energy
$E_0=\sum_k\frac{1}{2}\hbar\omega(k)$. The experimental
confirmation of a zero-point vacuum fluctuation was demonstrated
by the Casimir effect~\cite{Casimir}. A simple evaluation
of the vacuum density obtained from a summation of the zero-point
energy modes gives
\begin{equation}
\rho_{\rm vac}
=\frac{1}{(2\pi)^2}\int_0^{M_c}dkk^2(k^2+m^2)^{1/2}
\sim\frac{M^4_c}{16\pi^2},
\end{equation}
where $M_c$ is the cutoff. Already at the level of the standard
model, we get $\rho_{\rm vac}\sim (10^2\,{\rm GeV})^4$ which is
$55$ orders of magnitude larger than the bound (\ref{vacbound}).
To agree with the experimental bound (\ref{vacbound}), we would
have to invoke a very finely tuned cancellation of $\lambda_{\rm
vac}$ with the ``bare '' cosmological constant $\lambda$, which is
generally conceded to be theoretically unacceptable.

We can understand this result by using the language of Feynman
graphs. To avoid undue technical issues in FQFT, we shall consider
initially the basic lowest order vacuum fluctuation diagram computed
from the matrix element in flat Minkowski spacetime
\begin{equation}
M_{(2)}^{(0)}\sim
g^2\int d^4pd^4p'd^4k\delta(k+p-p')\delta(k+p-p')
$$ $$
\times\frac{1}{k^2+m^2}{\rm Tr}\biggl(\frac{i\gamma^\sigma
p_\sigma-m_f}{p^2+m_f^2}\gamma^\mu\frac{i\gamma^\sigma
p'_\sigma-m_f}{p^{'2}+m_f^2}\gamma_\mu\biggl)
$$ $$
\times\exp\biggl[-\biggl(\frac{p^2+m_f^2}{\Lambda_{\rm
SM}^2}\biggr) -\biggl(\frac{p'^2+m_f^2}{\Lambda_{\rm
SM}^2}\biggr) -\frac{k^2}{\Lambda_{\rm SM}^2}\biggr)\biggr],
\end{equation}
where $g$ is a coupling constant associated with the
standard model. We have considered a closed loop made of
a standard model fermion of mass $m_f$, an antifermion of the
same mass and an internal standard model boson propagator of mass
$m$; the scale $\Lambda_{\rm SM}\sim 10^2-10^3$ GeV. This leads
to the result
\begin{equation} M_{(2)}^{(0)}\sim
16\pi^4g^2\delta^4(a)\int_0^\infty dpp^3\int dp'p^{'3}
\biggl[\frac{-P^2+p^2+p^{'2}+4m_f^2}
{(P+a)(P-a)}\biggr]
$$ $$
\times\frac{1}{(p^2+m_f^2)(p^{'^2}+m_f^2)}
\exp\biggl[-\frac{(p^2+p^{'2}+2m_f^2)}{\Lambda_{\rm SM}^2}
-\frac{P^2}{\Lambda_{\rm SM}^2}\biggr],
\end{equation}
where $P=p-p'$ and $a$ is an infinitesimal constant which
formally regularizes the infinite volume factor $\delta^4(0)$. We
see that $\rho_{\rm vac}\sim M_{(2)}^{(0)}$ is finite and
$M_{(2)}^{(0)}\sim \Lambda_{\rm SM}^4$. To maintain gauge
invariance and unitarity in FQFT, we must add to this result the
contributions from the ghost diagram, the shadow diagram and the
measure diagram.

In flat Minkowski spacetime, the sum of all {\it disconnected}
vacuum diagrams $C=\sum_nM^{(0)}_n$ is a constant factor in the
scattering S-matrix $S'=SC$. Since the S-matrix is unitary
$\vert S'\vert^2=1$, then we must conclude that $\vert
C\vert^2=1$, and all the disconnected vacuum graphs can be
ignored. However, due to the equivalence principle {\it gravity
couples to all forms of energy}, including the vacuum energy
density $\rho_{\rm vac}$, so we can no longer ignore these
virtual quantum fluctuations in the presence of a non-zero gravitational
field.

Let us now consider the dominant contributions to the vacuum
density arising from the graviton loop corrections. As
explained above, we shall perform the calculations by expanding
about flat spacetime and trust that the results still hold for
an expansion about a curved metric background field, which is
strictly required for a non-zero cosmological constant. Since
the scales involved in the final answer, including the
predicted smallness of the cosmological constant, correspond to a
very small curvature of spacetime, we expect that our
approximation is justified.

We shall adopt a simple model consisting of a massive vector
meson $V_\mu$, which has the standard model
energy scale $\sim 10^2-10^3$ GeV. We have for the vector
field Lagrangian density
\begin{equation}
{\cal L}_V=-\frac{1}{4}(-{\bf g})^{-1/2}{\bf g}^{\mu\nu}
{\bf g}^{\alpha\beta}F_{\mu\alpha}
F_{\nu\beta}+m_V^2V_\mu V^\mu,
\end{equation} where
\begin{equation}
F_{\mu\nu}=\partial_\nu V_\mu-\partial_\mu V_\nu.
\end{equation}
We include in the Lagrangian density an additional piece
$-\frac{1}{2}(\partial_\mu V^\mu)^2$, and the vector field
propagator has the form
\begin{equation}
D^{\rm V}_{\mu\nu}
=\frac{\eta_{\mu\nu}}{p^2+m_V^2-i\epsilon}
\exp\biggl[-(p^2+m^2_V)/\Lambda_{\rm SM}^2\biggr]
$$ $$
=\eta_{\mu\nu}\int^\infty_1
\frac{d\tau}{\Lambda^2_{\rm SM}}\exp\biggl[-\tau (p^2+m_V^2)/\Lambda^2_{\rm SM}\biggr],
\end{equation}
while the shadow propagator is
\begin{equation}
D^{\rm shad\,V}_{\mu\nu}=
\frac{\eta_{\mu\nu}}{p^2+m^2_V}
\biggl[1-\exp\biggl[-(p^2+m^2_V)/\Lambda^2_{\rm SM}\biggr]\biggr]
$$ $$ =\eta_{\mu\nu}\int^1_0 \frac{d\tau}{\Lambda^2_{\rm SM}}
\exp\biggl[-\tau (p^2+m^2_V)/\Lambda^2_{\rm SM}\biggr].
\end{equation}

The graviton-V-V vertex in momentum space is given by
\begin{equation}
{\cal V}_{\alpha\beta\lambda\sigma}(p,q_1,q_2)
=\eta_{\lambda\sigma}
q_{1(\alpha}q_{2\beta)}-\eta_{\sigma(\beta}q_{1\alpha)}q_{\lambda}
-\eta_{\lambda(\alpha}q_{1_\sigma}q_{2\beta)}
$$ $$
+\eta_{\sigma(\beta}\eta_{\alpha)\lambda}q_1{\cdot q_2}
-\frac{1}{D-2}\eta_{\alpha\beta}(\eta_{\lambda\sigma}
q_1q_2-q_{1\sigma}q_{2\lambda}),
\end{equation}
where $q_1,q_2$ denote the momenta of the two $Vs$ connected to
the graviton with momentum $p$. We use the notation
$A_{(\alpha}B_{\beta)}=\frac{1}{2}(A_\alpha B_\beta+A_\beta
B_\alpha)$.

The lowest order correction to the graviton vacuum loop will have
the form
\begin{equation}
\label{PolV}
\Pi^V_{\mu\nu\rho\sigma}(p)
=-\kappa^2\exp\biggl(-p^2/\Lambda^2_G\biggr)\int d^4q
{\cal V}_{\mu\nu\lambda\alpha}(p,-q,q-p)
D^{V,W\,\lambda\delta}(-q) $$ $$
\times{\cal V}_{\rho\sigma\kappa\delta}(-p,p-q,q)
D^{V,W\,\alpha\kappa}(q-p).
\end{equation}

We obtain
\begin{equation}
\label{Ptensor}
\Pi^V_{\mu\nu\rho\sigma}(p)=-\kappa^2\exp\biggl(-p^2/\Lambda_G^2\biggr)
\int
\frac{d^4q}{(q^2+m_V^2)[(q-p)^2+m^2_V]}K_{\mu\nu\rho\sigma}(p,q)
$$ $$ \times\exp\biggl[-(q^2+m^2_V)/\Lambda^2_{\rm SM}\biggr]
\exp\biggl\{-[(q-p)^2+m^2_V]/\Lambda^2_{\rm SM}\biggr\},
\end{equation}
where in D-dimensions
\begin{equation}
K_{\mu\nu\rho\sigma}(p,q)=p_\alpha p_\beta p_\rho p_\sigma
+q_\alpha p_\beta p_\rho p_\sigma -q_\alpha q_\beta p_\rho
p_\sigma+(1-D)q_\alpha q_\beta q_\rho p_\sigma
$$ $$
-(1+D)p_\alpha q_\beta q_\rho q_\sigma+(D-1)p_\alpha q_\beta
p_\rho q_\sigma +Dq_\alpha q_\beta q_\rho q_\sigma.
\end{equation}
As usual, we must add to (\ref{PolV}) the contributions from the
fictitious ghost particle diagrams, the shadow field diagrams and
the invariant measure diagram.

We observe that from power counting of the momenta in the
integral (\ref{Ptensor}), we obtain
\begin{equation}
\Pi^V_{\mu\nu\rho\sigma}(p)\sim
\kappa^2\Lambda_{\rm SM}^4\exp\biggl(-p^2/\Lambda^2_G\biggr)
N_{\mu\nu\rho\sigma}(p^2)
$$ $$
\sim\frac{\Lambda_{\rm SM}^4}{M^2_{\rm PL}}
\exp\biggl(-p^2/\Lambda^2_G\biggr)N_{\mu\nu\rho\sigma}(p^2),
\end{equation}
where $N(p^2)$ is a finite remaining part of $\Pi^V(p)$. We
have as $p^2\rightarrow 0$:
\begin{equation}
{{N_{\mu}}^{\mu\sigma}}_\sigma(p^2)\sim p^4.
\end{equation}
Thus, $\Pi^V_{\mu\nu\rho\sigma}(p)$ vanishes at $p^2=0$
as it should because of gauge invariance and the massless
graviton.

For four-dimensional Euclidean momenta $p^2 \gg
\Lambda^2_G$, $\Pi^V_{\mu\nu\rho\sigma}(p)$ is exponentially
damped. At some value of the external graviton momentum $p$, when
$\Lambda^4_{\rm SM}$ could begin to become significant, the
exponential damping suppresses this contribution. If we
choose $\Lambda_G\leq 10^{-4}$ eV, then due to the damping of the
gravitational vacuum polarization loop graph in the Euclidean
limit $p^2\gg\Lambda_G^2$, {\it the cosmological constant
contribution is suppressed sufficiently to satisfy the bound
(\ref{vacbound})}, and it is protected from large unstable
radiative corrections. Thus, FQFT provides a solution to the
cosmological constant problem at the energy level of the
standard model and possible higher energy extensions of the
standard model. The universal fixed FQFT gravitational scale
$\Lambda_G$ corresponds to the fundamental length $\ell_G\leq 1$
cm at which virtual gravitational radiative corrections are cut
off.

We observe that the required suppression of the vacuum diagram
loop contribution to the cosmological constant, associated with
the vacuum energy momentum tensor at lowest order,
demands a low fundamental energy scale $\Lambda_G\leq
10^{-4}$ eV, which controls the quantum gravity loop
contributions. This is essentially because the external
graviton momenta are close to the mass shell, requiring a low
energy scale $\Lambda_G$. This seems at first sight a radical
suggestion that quantum gravity corrections are weak at energies
higher than $\leq 10^{-4}$ eV, but this is clearly not in
contradiction with any known gravitational experiment. Indeed, as
has been stressed in recent work on large higher dimensions,
there is no experimental knowlege of gravitational forces below 1
mm. In fact , we have no experimental knowledge at present about
the strength of graviton radiative corrections. The standard
model experimental agreement is achieved for standard model
particle states close to the mass shell. However, we expect that
the dominant contributions to the vacuum density arise from
standard model states far off the mass shell. In our perturbative
quantum gravity theory, the tree graphs involving gravitons are
identical to the tree graphs in local point graviton perturbation
theory, retaining classical, causal GR and Newtonian gravity. In
particular, {\it we do not decrease the strength of the
classical, large distance gravity force.}

In order to solve the severe cosmological constant hierarchy
problem, we have been led to the surprising conclusion that,
in contrast to the conventional folklore, quantum gravity
corrections to the classical GR theory are negligible at
energies above $\leq 10^{-4}$ eV, a result that will continue
to persist if our perturbative calculations can be extrapolated
to near the Planck energy scale $\sim 10^{19}$ GeV. Since the
cosmological constant problem already results in a severe crisis
at the energies of the standard model, our quantum gravity
resolution based on perturbation theory can resolve the
crisis at the standard model energy scale and well beyond this
energy scale.

\section{\bf Conclusions}

The ultraviolet finiteness of perturbative quantum field
theory in four-dimensions is achieved by applying
the FQFT formalism. The nonlocal quantum loop interactions
reflect the quantum, non-point-like nature of the field theory, although
we do not specify the nature of the extended object that
describes a particle. Thus, as with string theories, the
point-like nature of particles is ``fuzzy'' in FQFT for energies
greater than the scale $\Lambda$. One of the features of
superstrings is that they provide a mathematically consistent
theory of quantum gravity, which is ultraviolet finite and
unitary. FQFT focuses on the basic mechanism behind string
theory's finite ultraviolet behavior by invoking a suppression of
bad vertex behavior at high energies, without compromising
perturbative unitarity and gauge invariance. FQFT provides a
mathematically consistent theory of quantum gravity at the
perturbative level. If we choose $\Lambda_G\leq 10^{-4}$ eV, then
quantum radiative corrections to the classical tree graph gravity
theory are perturbatively negligible to all energies greater than
$\Lambda_G$, provided that the perturbative regime is valid.

The important gauge hierarchy problem, associated with the Higgs sector, is
solved by the exponential damping of the Higgs self-energy in the Euclidean
$p^2$ domain for $p^2 \gg \Lambda_H^2$, and for a $\Lambda_H$
scale in the electroweak range $\sim 10^2-10^3$ GeV. A damping
of the vacuum polarization loop contributions to the vacuum
energy density-gravity coupling at lowest order can resolve the
cosmological constant hierarchy problem, if the gravity loop
scale $\Lambda_G\leq 10^{-4}$ eV, by suppressing virtual
gravitational radiative corrections above the energy scale
$\Lambda_G$.

We must still set the physical scale $\Lambda_{\rm YM}$, which
controls the size of radiative loop corrections in the
Yang-Mills sector of FQFT. We expect this scale to
be much larger than the electroweak scale $\sim 10^2-10^3$ Gev,
and it could be as large as grand unification theory (GUT) scales
$\sim 10^{16}$ Gev, allowing for possible GUT unification
schemes.

Recently, new supernovae data have strongly indicated a
cosmic acceleration of the present
universe~\cite{Perlmutter}. This has brought the status of
the cosmological constant back into prominence, since one
possible explanation for this acceleration of the expansion of the
universe is that the cosmological constant is non-zero but very
small. We can, of course, accomodate a small non-zero
cosmological constant by choosing carefully the gravity scale
$\Lambda_G$. Indeed, this new observational data can be viewed as
a means of determining the size of $\Lambda_G$.

Our quantum field theory formalism has helped to
resolve two critical hierarchy problems in modern physics,
given two parameters $\Lambda_H\sim 10^2-10^3$ GeV and
$\Lambda_G\leq 10^{-4}$ GeV. These parameters will hopefully be
explained by a more fundamental non-perturbative theory.
\vskip 0.2
true in {\bf Acknowledgments}
\vskip 0.2 true in
I thank Michael Clayton, John Donoghue, Holger Nielsen, Bob
Holdom, George Leibbrandt, Michael Luke, Anton Kapustin, Pierre
Savaria, Raman Sundrum and Gerard 't Hooft for helpful and
stimulating discussions. This work was supported by the Natural
Sciences and Engineering Research Council of Canada.
\vskip 0.5
true in

\end{document}